\documentclass[aps, prl, 10pt, twocolumn, superscriptaddress, amsmath, amssymb]{revtex4-2}

\usepackage[T1]{fontenc}		
\usepackage{graphicx}     	
\usepackage{hyperref}				
\usepackage{lmodern}				
\usepackage{physics}				
\usepackage{upgreek}				
\usepackage{natbib}					
\usepackage{array}					
\usepackage{bm}							
\usepackage{nameref}

\hypersetup{
    colorlinks=true,
    linkcolor=blue,
    filecolor=blue,      
    urlcolor=blue,
    citecolor=blue
}

\begin{document}

\title{All-Optical Nuclear Quantum Sensing using Nitrogen-Vacancy Centers in Diamond}

\author{B. Bürgler} \thanks{These authors contributed equally.}
\affiliation{Department of Physics, University of Basel, Klingelbergstrasse 82, Basel CH-4056, Switzerland}

\author{T. F. Sjolander} \thanks{These authors contributed equally.}
\affiliation{Department of Physics, University of Basel, Klingelbergstrasse 82, Basel CH-4056, Switzerland}

\author{O. Brinza}
\affiliation{Laboratoire des Sciences des Procédés et des Matériaux, LSPM, CNRS-UPR 3407, Université Sorbonne Paris Nord, 99 Avenue JB Clément, Villetaneuse 93430, France}

\author{A. Tallaire}
\affiliation{Laboratoire des Sciences des Procédés et des Matériaux, LSPM, CNRS-UPR 3407, Université Sorbonne Paris Nord, 99 Avenue JB Clément, Villetaneuse 93430, France}
\affiliation{Institut de Recherche de Chimie Paris, CNRS, Chimie ParisTech, Université PSL, 11 rue Pierre et Marie Curie, 75005 Paris, France}

\author{J. Achard}
\affiliation{Laboratoire des Sciences des Procédés et des Matériaux, LSPM, CNRS-UPR 3407, Université Sorbonne Paris Nord, 99 Avenue JB Clément, Villetaneuse 93430, France}

\author{P. Maletinsky} \email{patrick.maletinsky@unibas.ch}
\affiliation{Department of Physics, University of Basel, Klingelbergstrasse 82, Basel CH-4056, Switzerland}

\date{\today}

\begin{abstract}
Solid state spins have demonstrated significant potential in quantum sensing with applications including fundamental science, medical diagnostics and navigation. 
The quantum sensing schemes showing best performance under ambient conditions all utilize microwave or radio-frequency driving, which poses a significant limitation for miniaturization, energy-efficiency and non-invasiveness of quantum sensors. 
We overcome this limitation by demonstrating a purely optical approach to coherent quantum sensing. 
Our scheme involves the $^{15}$N nuclear spin of the Nitrogen-Vacancy (NV) center in diamond as a sensing resource, and exploits NV spin dynamics in oblique magnetic fields near the NV’s excited state level anti-crossing to optically pump the nuclear spin into a quantum superposition state. 
We demonstrate all-optical free-induction decay measurements - the key protocol for low-frequency quantum sensing - both on single spins and spin ensembles. 
Our results pave the way for highly compact quantum sensors to be employed for magnetometry or gyroscopy applications in challenging environments.
\end{abstract}

\maketitle

     \section{Introduction}

Spin based quantum sensors can be employed to measure a wide range of relevant physical quantities, including magnetic\,\cite{Taylor2008a} or electric fields\,\cite{Dolde2011a}, temperature\,\cite{Acosta2010a}, or rotary motion\,\cite{Ledbetter2012a}.
This abundance of potential observables, combined with their high sensitivity at the nano-scale makes quantum sensors highly interesting for many fields of application, such as life sciences\,\cite{Barry2016a}, geological sciences\,\cite{Glenn2017a,Levine2019a}, navigation\,\cite{Phillips2020a} and material sciences\,\cite{Casola2018a}.

Nitrogen-Vacancy (NV) centers in diamond (Fig.\,\ref{Fig1}a) are a particularly promising platform for such spin based quantum sensing applications, because they host a single electron spin\,\cite{Doherty2013a} with long coherence times\,\cite{Balasubramanian2009a, BarGill2013a} even at room temperature\,\cite{Kennedy2003a}.  
Upon optical excitation with green light, the NV center emits spin-dependent red photoluminescence (PL)\,\cite{Gruber1997a}, which enables all-optical electron spin readout.  
At the same time, such optical excitation pumps the NV electron spin\,\cite{Harrison2004a, Tetienne2012a} into a specific spin eigenstate, enabling all-optical spin initialization.  
Time-varying (AC) driving fields, mostly in the microwave (MW) or radio-frequency (RF) domain can then be used to coherently control the spin, and create superposition states for sensing.
This combination of optical initialization, readout, and coherent spin manipulation by AC driving fields form the basis of almost all established spin based approaches to sensing\,\cite{Degen2017a}.

The NV electron spin is inherently coupled to the nuclear spin of its Nitrogen atom -- a spin which exhibits significantly longer coherence times compared to the NV electron spin\,\cite{Waldherr2012a, Pfender2017a} and therefore provides another interesting resource for quantum technology applications.  
Specifically, nuclear spins have been exploited as a quantum register for quantum communication\,\cite{Pompili2021a} and enhanced spin readout techniques\,\cite{Lovchinsky2016a}, but they also offer interesting opportunities for sensing, be it for magnetometry\,\cite{Waldherr2012a, Chen2013a, Jakobi2017a, Sahin2022a} or for gyroscopy\,\cite{Ajoy2012a, Jarmola2021a}.

Many Nitrogen spin based quantum sensing schemes in diamond rely on the resonant coupling of the NV spin and the nuclear Nitrogen spin at a magnetic field of about $500~$G\,\cite{Wood2016a,Jarmola2020a}, where spin flip-flop processes occur in the NV's orbital excited state at the excited state level anti-crossing (ESLAC).  
It has been shown that optical pumping in the vicinity of the ESLAC results in nuclear spin hyperpolarization\,\cite{Jacques2009a,Steiner2010a}, and that -- by virtue of the same mechanism -- the NV center shows a nuclear spin-state dependent rate of transient PL\,\cite{Jarmola2020a}.  
As a result, optical pumping close to the ESLAC enables both all-optical readout of the nuclear spin state and initialization into a nuclear spin eigenstate, which, together with RF driving, forms the basis for nuclear spin based sensing schemes\,\cite{Jarmola2020a}.

However, the ubiquitous need for AC coherent driving in spin based quantum sensing is a severe limitation for many applications.
Specifically, such AC driving fields can adversely affect investigated samples and for integrated or portable sensing devices, their delivery increases power-consumption and overall complexity, and thereby size and cost of the system.
Recent experiments have demonstrated microwave-free NV magnetometry schemes that are based on sharp changes in  NV PL at level anti-crossings, that occur at specific magnetic fields\,\cite{Wickenbrock2016a, Zheng2020a}.
While avoiding the need for MW or RF delivery, these approaches do not exploit quantum coherence (and are therefore limited in sensitivity) and are furthermore highly vulnerable to background drifts in NV PL.

Here, we present a novel method for coherent, microwave-free quantum sensing using the $^{15}$N nuclear spin of the NV center in diamond.  
Our approach is based on optical driving of the NV center near the ESLAC in the presence of a small, static magnetic field transverse to the NV symmetry axis denoted by the unit vector $\bm{e}_z$ (see Fig.\,\ref{Fig1}a).
As we will show, such a small transverse magnetic field component has a striking effect, in that optical pumping prepares the $^{15}$N nuclear spin in a coherent superposition state within the NV's ground state spin manifold.
Following optical pumping, this initialization leads to Larmor precession of the nuclear spin about an effective magnetic field; a precession we directly monitor via nuclear spin-state dependent PL\,\cite{Jarmola2020a}. 
Figure\,\ref{Fig1}b shows an example of such an all-optical nuclear free induction~(FID) measurement, obtained using the pulse sequence depicted in Fig.\,\ref{Fig1}c. 
Data for this work were recorded on a home-built confocal optical microscope (see methods) with magnetic field control; here at a magnetic field of strength $|\bm{B}_{\rm ext}| = 540~$\text{G}, tilted by $\Phi = 1^\circ$ away from $\bm{e}_z$.

The negatively charged NV center possesses an electron spin $S=1$ quantized along the NV symmetry axis~$\bm{e}_z$.
For NV centers formed by $^{15}$N (denoted as ``$^{15}$NV'' in the following), the Nitrogen nucleus exhibits a spin $I=1/2$. 
The Hamiltonian for the orbital ground~(gs) and excited state~(es) of such a $^{15}$NV can be expressed as
\begin{equation}
\frac{\hat{H}^{\rm gs, es}}{h} = D_0^{\rm gs, es}\hat{S}_z^2 + \hat{\bm{S}}\cdot\bm{A}^{\rm gs, es}\cdot\hat{\bm{I}} + \gamma_{S} \bm{B}_{\rm ext}\cdot\hat{\bm{S}} + \gamma_{I}\bm{B}_{\rm ext}\cdot\hat{\bm{I}}\,,
\end{equation}
where $\hat{\bm{S}}$ and $\hat{\bm{I}}$ are the NV electron and nuclear spin operators, $\gamma_{S}=2.8~$MHz/G and $\gamma_{I}=431.7~$Hz/G are the respective gyromagnetic ratios, $D_0^{\rm gs} = 2.87~$GHz and $D_0^{\rm es} = 1.42~$GHz are the zero-field splittings, and $\bm{B}_{\rm ext}$ is the applied magnetic field.
The hyperfine coupling tensor $\bm{A}^{\rm gs, es}$ has two independent components $A_\parallel^{\rm gs} = 3.03~$MHz and $A_\perp^{\rm gs} = 3.65~$MHz for the ground state; and $A_\parallel^{\rm es}=-57.8~$MHz, $A_\perp^{\rm es} = -39.2~$MHz for the excited state\,\cite{Gali2009a, Felton2009a}.
This Hamiltonian is conveniently expressed in a basis of spin eigenstates $\lbrace \ket{m_S,m_I} \rbrace$, where $m_S$ and $m_I$ are the magnetic quantum numbers associated with $\hat{S}_z$ and $\hat{I}_z$. 
The coherent FID dynamics that are studied in this work (Fig.\,\ref{Fig1}b) can be completely encompassed by a reduced subspace spanned by $\lbrace \ket{0,-1/2},\ket{0,+1/2} \rbrace$.

     \section{Results}

\begin{figure}[t]
\includegraphics{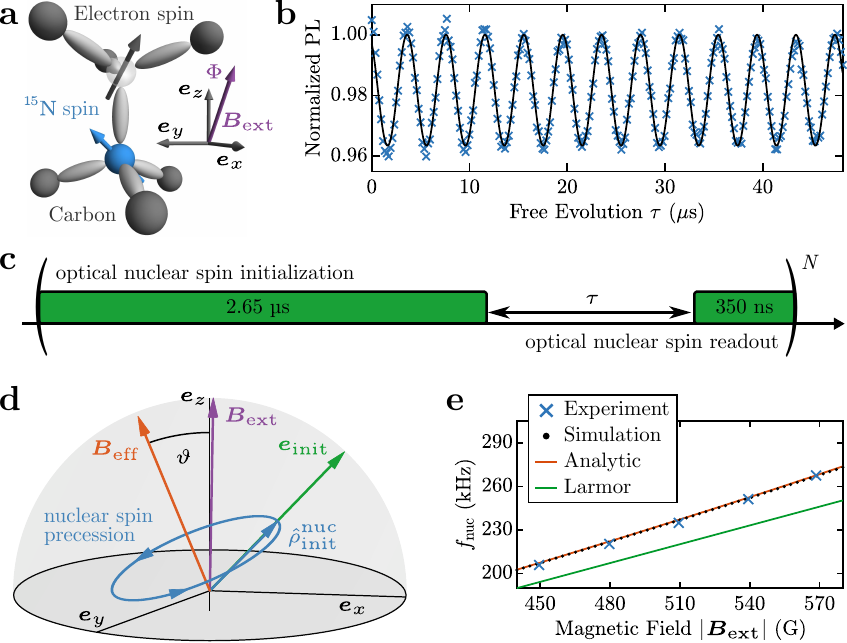}
\caption{\label{Fig1} 
\textbf{a}~Crystal structure of the Nitrogen-Vacancy center with illustration of its associated spins and coordinate axes. 
\textbf{b}~All-optical nuclear spin precession of the $^{15}$N Nuclear spin observed at a magnetic field of $|\bm{B}|= 540~$G tilted away from the NV symmetry axis by $\Phi=1^\circ$. 
Fitting of a harmonic function (black) yields a precession frequency $251.18\pm0.12~$kHz.  
\textbf{c}~Pulse sequence employed for \textbf{b}, consisting of a $3~$\textmu s green laser pulse separated by a variable delay $\tau$.  
The first $350~$ns of each green pulse are utilized for optical nuclear spin readout, while the remainder of the pulse reinitializes the spin system.
\textbf{d}~Quantitative Bloch-Sphere representation of the $^{15}$N spin in the $\ket{m_s=0}$ ground state manifold.  
For a magnetic field $\bm{B}_{\rm ext}$ tilted from the NV symmetry axis by the angle $\Phi=1^\circ$, optical pumping initializes the nuclear spin into $\hat{\rho}_{\rm init}^{\rm nuc}$. The nuclear spin subsequently precesses around an effective magnetic field $\bm{B}_{\rm eff}$.
The measurement axis for all-optical readout of this precession is given by $\bm{e}_{\rm init}$.
\textbf{e}~Experimentally observed precession frequency (blue crosses) at different values of $|\bm{B}_{\rm ext}|$ and fixed $\Phi=1^\circ$, together with analytic (solid orange line) and numerical predictions (black dots).}
\end{figure}

\textbf{Derivation of an effective Hamiltonian}. In the following we explain the nuclear precession data presented in Fig.\,\ref{Fig1}b by discussing how the presence of the transverse magnetic field component $B_\perp$ significantly affects the system's optical pumping and subsequent FID dynamics.
We start by calculating an effective Hamiltonian for the $^{15}$N spin in the $m_S=0$ ground state subspace, $\lbrace \ket{0,-1/2},\ket{0,+1/2} \rbrace$, using Van Vleck perturbation theory\,\cite{CohenTannoudji1998a} (see the supplementary information, Section\,\hyperref[SI1]{I}, for a detailed derivation). 
Without loss of generality, we set the transverse magnetic field to point along the unit vector $\bm{e}_x$. 
We then obtain the effective Hamiltonian
\begin{gather}
\frac{H^{m_S=0}_{\rm{eff}}}{h} = \frac{1}{2}\begin{bmatrix}\gamma_I B_z + \nu_z &\hspace{0.2cm} \gamma_I B_\perp + \nu_\perp \vspace{0.2cm}\\ \gamma_I B_\perp + \nu_\perp &\hspace{0.2cm} - \gamma_I B_z - \nu_z \end{bmatrix},
\label{eq:Heff}
\end{gather}
where
\begin{equation}
\nu_z = \frac{\gamma_SB_z(A_\perp^{\rm gs})^2}{(D_0^{\rm gs})^2-(\gamma_SB_z)^2}
\end{equation}
denotes the correction to the diagonal elements caused by mixing between states of different $m_S$, and 
\begin{equation}\label{nu_transverse}
\nu_\perp = \frac{-2\,\gamma_SB_\perp A_\perp^{\rm gs}D_0^{\rm gs}}{(D_0^{\rm gs})^2-(\gamma_SB_z)^2}
\end{equation}
is the corresponding correction to the off-diagonal elements.
We note that such an effective hyperfine Hamiltonian for $^{15}$NV's has been discussed earlier as a perturbation in $B_\perp$ in the limit $B_z\ll D_0^{\rm gs}$\,\cite{Childress2006a}, or as a perturbation in $A_\perp^{\rm gs}$\,\cite{Jarmola2020a}, but never as a perturbation in both simultaneously as we present it here.
Additionally, exact analytic expressions for $\nu_\perp$ have previously been derived in\,\cite{Chen2015a}.

Hamiltonian $H^{m_S=0}_{\rm{eff}}$ yields that the $^{15}$N nuclear spin in the NV ground state is quantized along an effective magnetic field $\bm{B}_{\rm{eff}}$. 
Diagonalization of $H^{m_S=0}_{\rm{eff}}$ yields
\begin{equation}
\gamma_I|\bm{B}_{\rm{eff}}| = \sqrt{\left(\gamma_I B_z + \nu_z\right)^2 + \left(\gamma_I B_\perp + \nu_\perp\right)^2} =: f_{\rm nuc}\,,
\label{Beff}
\end{equation}
where $f_{\rm nuc}$ is the expected nuclear spin precession frequency. 
Importantly, $\bm{B}_{\rm{eff}}$ is neither aligned with $\bm{B}_{\rm ext}$ nor with the NV symmetry axis.
Instead, it lies in the plane spanned by $\bm{B}_{\rm{eff}}$ and $\bm{B}_{\rm ext}$, and is tilted away from $\bm{e}_z$ by an angle $\vartheta = \tan^{-1}\left[(\gamma_I B_\perp + \nu_\perp) / (\gamma_I B_z + \nu_z)\right]$. 
Interestingly, $\vartheta$ is significantly larger than the misalignment angle $\Phi$ between $\bm{B}_{\rm ext}$ and $\bm{e}_z$, and $\vartheta$ has a sign opposite to $\Phi$ due to the negative sign of $\gamma_I$ (c.f. Fig.\,\ref{Fig1}d). Finally, note that according Eq.\,\eqref{nu_transverse}, $\nu_\perp = 0$ when $B_\perp=0$, which in turns causes $\vartheta=0$. 
In this case, both $\bm{B}_{\rm eff}$ and $\bm{B}_{\rm ext}$ are aligned with the NV symmetry axis.

\textbf{Analysis of $^{15}$N spin dynamics}. We now discuss how the presence of $B_\perp$ affects the $^{15}$N nuclear spin dynamics and enables all-optical initialization into a nuclear spin superposition state.
The use of $^{15}$N is key to this, since it does not have a quadrupolar spin splitting which would prevent any nuclear Larmor precession.
It is clear from Eq.\,\eqref{nu_transverse} that when $B_\perp\neq0$, the $^{15}$N nuclear quantization axis depends sensitively on the hyperfine coupling parameter~$A_\perp$, and the splitting of the involved spin-levels.
Therefore, the ground and excited state nuclear spin quantization axes are in general different, because the hyperfine parameters differ in both magnitude and sign between the two cases.
This difference in $^{15}$N quantization axes results in optical pumping of the nuclear spin into a state that does not correspond to an eigenstate of the effective ground state Hamiltonian $H^{m_S=0}_{\rm{eff}}$. 
To be specific, optical pumping accumulates NV excited state population in the eigenstate with the largest $m_S=0$ character (i.e. the state $\ket{\psi}$ for which $\bra{\psi}\hat{\bm{S}}_z/\hbar\ket{\psi}$ is closest to zero), since this state has the lowest probability of shelving into the NV's singlet state -- we denote this state as $\ket{\tilde{0}_{\rm es}}$.
By the same argument, $\ket{\tilde{0}_{\rm es}}$ is also the ``brightest'' state in that it yields the largest rate of emission of NV fluorescence photons.  
For state $\ket{\tilde{0}_{\rm es}}$, the expectation value of the nuclear spin lies along the vector $\bm{e}_{\rm init}=\bra{\tilde{0}_{\rm es}}\hat{\bm{I}}/\hbar\ket{\tilde{0}_{\rm es}}$.
Vector $\bm{e}_{\rm init}$ therefore defines both the direction along which the nuclear spin is initialized under green illumination, as well as the measurement axis for optical readout of the nuclear spin.
For $B_\perp\neq0$, $\bm{e}_{\rm init}$ is not collinear with $\bm{B}_{\rm{eff}}$, and thus optical pumping will initialize the $^{15}$N nuclear spin in a state that is not an eigenstate of $H^{m_S=0}_{\rm{eff}}$.
Disengaging green laser excitation after optical pumping will therefore result in precession of the $^{15}$N nuclear spin around $\bm{B}_{\rm{eff}}$.
Finally, note that for NV centers formed with $^{14}$N, no Larmor precession occurs because its quadrupolar splitting locks $\bm{B}_{\rm{eff}}$ onto the NV axis.

\textbf{Comparison with numerical results}. To further verify this picture, we developed a numerical model that simulates the dynamics of the $^{15}$NV system during and after optical pumping.
The model is based on classical rate equations for the optical pumping process\,\cite{Tetienne2012a}, coupled with master equations describing the quantum-mechanical evolution of the system's density matrix within each relevant orbital manifold: the orbital ground and excited states as well as the singlet state (see further details in the supplementary information, Section\,\hyperref[SI21]{II.1}).

In Fig.\,\ref{Fig1}d we summarize the numerical and theoretical results in a Bloch sphere representation of the $^{15}$N nuclear spin dynamics for the same magnetic field that was used to obtain the experimental results in Fig.\,\ref{Fig1}b.
The effective field $\bm{B}_{\rm eff}$ is calculated numerically through exact diagonalisation of the ground state Hamiltonian $\hat{H}^{\rm gs}$, and $\bm{e}_{\rm init}$ is calculated via diagonalization of the excited state Hamiltonian $\hat{H}^{\rm es}$.
The nuclear spin density matrix $\hat{\rho}_{\rm init}^{\rm nuc}$ following optical pumping is obtained by propagating the system density matrix $\hat{\rho}$ for $3$~\textmu s of laser excitation, followed by a $50~$ns dark time (to let the system relax fully to the ground state), and by subsequently taking the trace over the NV electron spin degrees of freedom.
The nuclear spin precession dynamics is then described by propagating $\hat{\rho}$ under the influence of $\hat{H}^{\rm gs}$.

We make two observations that underline the excellent agreement of this numerical model with our analytical discussion. 
First, the orientation of $\bm{B}_{\rm eff}$ obtained from numeric diagonalization of the full ground state Hamiltonian $\hat{H}^{\rm gs}$ shows perfect agreement with the analytical prediction from Eq.\,\eqref{eq:Heff} (see further details in the supplementary information, Section\,\hyperref[SI23]{II.3}).
Second, the initial nuclear spin direction Tr$(\hat{\bm{I}}\hat{\rho}_{\rm init}^{\rm nuc})$ obtained from our numerical model is perfectly collinear with $\bm{e}_{\rm init}$, as long as we set the intersystem crossing rate for the $m_S=0$ states to zero (see the supplementary information, Section\,\hyperref[SI23]{II.3}).
Both observations strongly support the validity of our model and its applicability to quantitatively describe our data.

\textbf{All-optical nuclear $^{15}$N precession} The presented theory framework allows us to further analyze the data presented in Fig.\,\ref{Fig1}b.
The observed FID oscillation frequency was determined by least-squares fitting to $f_{\rm nuc}=251.18\pm0.12~$kHz, in good agreement with Eq.\,\eqref{Beff}, which yields $f_{\rm nuc} = 252.71~$kHz for the experimental conditions $|\bm{B}_{\rm ext}|=540$\,G and $B_\perp=10.6$\,G.
The small remaining discrepancy can be assigned to uncertainties in controlling the tilt angle $\Phi$, and determining the exact field components $B_\perp$ and $B_\parallel$.
To demonstrate that the observed oscillations indeed originate from nuclear spin precession, we repeated the same experiment at fixed angle $\Phi$, while varying $|\bm{B}_{\rm ext}|$.  
Figure\,\ref{Fig1}e shows the resulting, near-linear dependance of $f_{\rm nuc}$ on $|\bm{B}_{\rm ext}|$ and the excellent agreement with the predictions of both Eq.\,\eqref{Beff} and the numeric model.
The enhancement of $f_{\rm nuc}$ over the bare Larmor frequency results from the terms $\nu_z$ and $\nu_\perp$ in Eq.\,\eqref{Beff}.

\begin{figure}[t]
\includegraphics{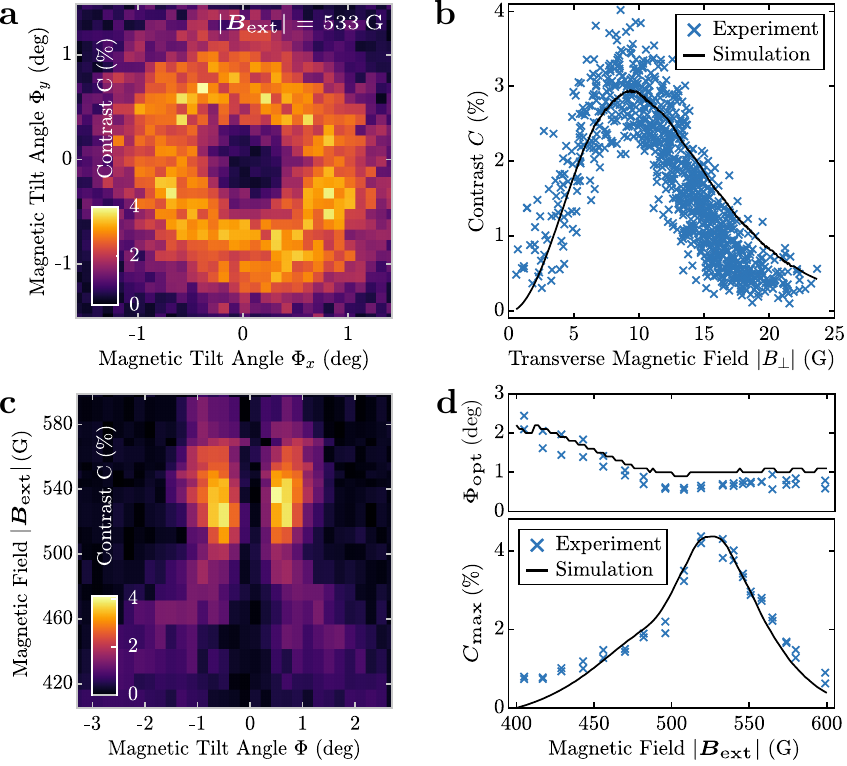}
\caption{\label{Fig2} 
\textbf{a}~Measured free induction decay (FID) contrast of a single NV as a function of magnetic field orientation, for a fixed total field of $|\bm{B}_{\rm ext}| = 533~$G.
\textbf{b}~Same data as in \textbf{a}, but plotted against total transverse magnetic field $B_\perp$. The black line is the prediction of our numerical model, which is normalized to the mean of the data at $\Phi_{\rm opt}\pm0.03^\circ$.
\textbf{c}~Nuclear FID contrast as a function of magnetic tilt angle $\Phi$ and total magnetic field $|\bm{B}_{\rm ext}|$.  
\textbf{d}~Maximal observed contrast $C_{\rm max}$ and corresponding tilt angle $\Phi_{\rm opt}$ for each value of $|\bm{B}_{\rm ext}|$.  
The prediction of our numerical model is shown in black, which is normalized to the maximal data point.}
\end{figure}

\textbf{Angle ($\Phi$) and field ($|\bm{B}_{\rm ext}|$) dependance of the all-optical $^{15}$N FID signal}.  
The requirement of applying a transverse magnetic field $B_\perp$ to obtain an observable, all-optical $^{15}$N FID signal motivates the question of how the FID readout contrast $C$ depends on both $\Phi$ and $|\bm{B}_{\rm ext}|$.  
Figure\,\ref{Fig2}a shows single NV data, where we determined $C$ as a function of $\Phi$ for a fixed field $|\bm{B}_{\rm ext}|=533~$G. 
We determined $C$ from the Fourier space amplitude of individual FID curves, for varying field misalignment angles $\Phi_x$ and $\Phi_y$, applied in the $x$-$z$ and $y$-$z$-planes, respectively.  
For the small angles we investigated, $\Phi\approx (\Phi_x^2+\Phi_y^2)^{1/2}$.
As expected, when $\Phi=0$, no nuclear FID contrast is observed, because in this case $\bm{B}_{\rm eff}$ and $\bm{e}_{\rm init}$ are both collinear with $\bm{e}_z$ such that the nuclear spin is optically pumped into the non-precessing eigenstate $\ket{0,+1/2}$ of $\hat{H}^{\rm gs}$.   
Upon increasing $\Phi$, $\bm{B}_{\rm eff}$ and $\bm{e}_{\rm init}$ both tilt away from $\bm{e}_z$ in different directions, resulting in nuclear precession of increasing contrast $C$. 
At the same time, increasing $\Phi$ (i.e. $B_\perp$) tends to reduce the nuclear hyperpolarization efficiency\,\cite{Jacques2009a, Jarmola2020a} and the NV optical spin readout contrast\,\cite{Tetienne2012a}, both of which reduce $C$.
Overall, these counteracting effects imply that there is an optimal tilt angle $\Phi_{\rm opt}$ which maximizes $C$ by balancing the magnitude of the nuclear spin coherences with nuclear spin readout efficiency.  
We call this maximized contrast $C_{\rm max}$.

To determine $\Phi_{\rm opt}$ and $C_{\rm max}$, we show in Fig.\,\ref{Fig2}b the data from Fig.\,\ref{Fig2}a as a function of transverse field $B_\perp$, where for each data point, we determined $B_\perp$ from the NV's full optically detected magnetic resonance spectrum. 
Figure\,\ref{Fig2}b reveals a clear maximum in $C$ at $B_{\perp}\approx8.6~$G, which corresponds to $\Phi_{\rm opt}\approx 0.86^\circ$ (see further details in the supplementary information, Section\,\hyperref[SI3]{III}).
These results are in good agreement with the predictions of our numerical model (black curve in in Fig.\,\ref{Fig2}b).
The quality of the simulation depends sensitively on the NV intersystem crossing rates, all of which were kept fixed to literature values\,\cite{Tetienne2012a} in our calculations (see further details in the supplementary information, Section\,\hyperref[SI23]{II.3}).
We assign remaining discrepancies between data and simulations to uncertainties on optical transition rates employed in the model.

Interestingly, we find that our all-optical $^{15}$N nuclear FID protocol is relatively resilient to deviations of $\bm{B}_{\rm ext}$ away from ideal ESLAC conditions.
For this, we investigated the dependance of the contrast $C$ on the applied magnetic field $|\bm{B}_{\rm ext}|$ and tilt angle $\Phi_x$, where for each data point, we ensured that $\Phi_y=0$ is maintained to within experimental accuracy.  
The resulting data show a nontrivial dependance of $C$ on $|\bm{B}_{\rm ext}|$ and $\Phi_x$ (Fig.\,\ref{Fig2}c), and in particular reveal that both $\Phi_{\rm opt}$ and $C_{\rm max}$ change with $|\bm{B}_{\rm ext}|$ (Fig.\,\ref{Fig2}d).
These dependencies are qualitatively captured by our numerical model. 
We find a global maximum of $C_{\rm max}\approx 4.2~\%$ for $|\bm{B}_{\rm ext}|=533~$G, and a drop of $C$ over a full-width at half maximum (FWHM) range of $\sim50$~G.

\begin{figure}[t]
\includegraphics{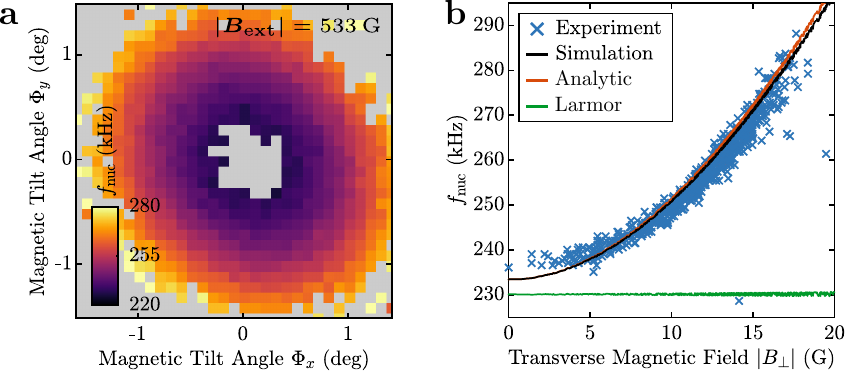}
\caption{\label{Fig3} 
\textbf{a}~$^{15}$N nuclear precession frequency $f_{\rm nuc}$ corresponding to the data shown in Fig.\,\ref{Fig2}a. Only pixels for which $C>1\%$ are shown.  
\textbf{b}~Same data as in \textbf{a}, plotted against total transverse magnetic field $B_\perp$, together with the numerical model prediction (black), the prediction from Eq.\,\eqref{Beff} (orange), and the bare Larmor frequency $\gamma_{I}B_z$ (green). 
}
\end{figure}

\textbf{Nuclear spin precession frequency}.
Further, we investigate the dependance of $f_{\rm nuc}$ on the magnetic field tilt angles $\Phi_x$ and $\Phi_y$.
For this, we determine $f_{\rm nuc}$ by Fourier analysis of the FID data for each data point sampled in Fig.\,\ref{Fig2}a.
The result is shown in Fig.\,\ref{Fig3}a, along with the corresponding plot of the same data as a function of $|B_\perp|$ in Fig.\,\ref{Fig3}b.
The precession frequency $f_{\rm nuc}$ increases with $B_\perp$ in a way that is excellently described by both Eq.\,\eqref{Beff} and our numerical model. 
We again assign small discrepancies between measured and predicted values of $f_{\rm nuc}$ to experimental uncertainties in determining $\bm{B}_{\rm ext}$.

\begin{figure*}[t]
\includegraphics{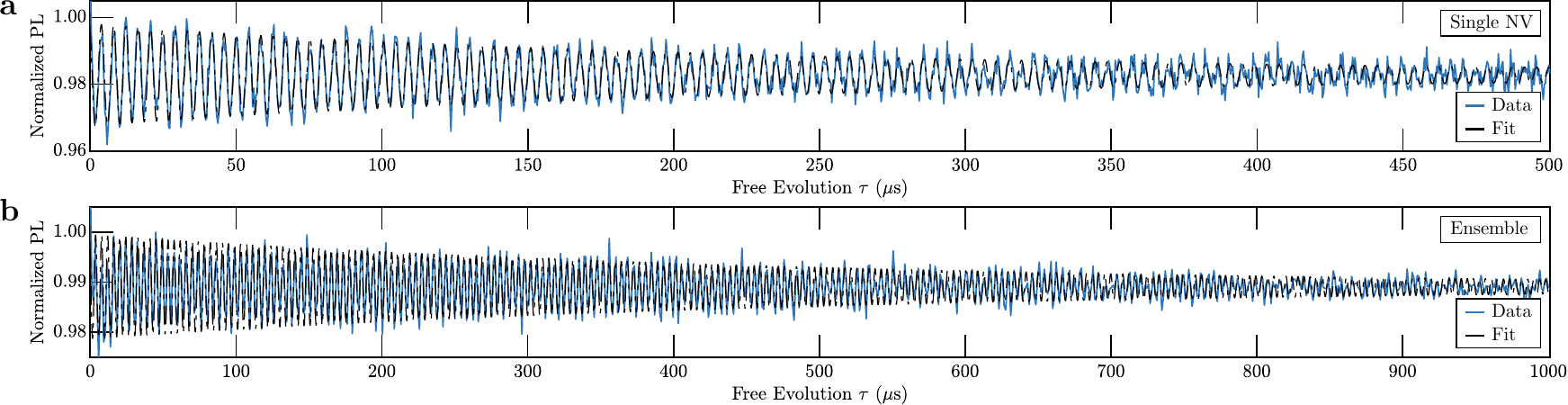}
\caption{\label{Fig4} All-Optical nuclear spin precession of \textbf{a}, a single NV in a diamond nanopillar, and \textbf{b}, an ensemble of NVs in bulk diamond, both measured at $|\bm{B}_{\rm ext}|=533~$G with tilt angles $\Phi=0.65^\circ$ and $\Phi=0.8^\circ$, for \textbf{a} and \textbf{b} respectively.
Each data set is fitted with a damped harmonic function to determine the nuclear spin coherence time $T_2^*$, yielding $T_2^*=248.1\pm 12.4~$\textmu s and $T_2^*=508.5\pm 17.4~$\textmu s, for \textbf{a} and \textbf{b} respectively.}
\end{figure*}

\textbf{Nuclear coherence time}. The ease of use of our all-optical $^{15}$N FID experiments enables facile assessment of the $^{15}$N inhomogeneous nuclear spin coherence time $T_2^*$. 
To determine $T_2^*$, we extend the measurement pulse sequence shown in Fig.\,\ref{Fig1}c to longer FID evolution times $\tau$, to resolve the full decay of the signal (Fig.\,\ref{Fig4}a).
Fitting an exponentially decaying harmonic function to the data yields $C=3.09\pm0.11~$\% and $T_2^* = 248.1\pm 12.4~$\textmu s for the single NV center under investigation.
This decoherence time is somewhat shorter than previously reported values\,\cite{Sangtawesin2016a}, but consistent with the rather short NV electron spin relaxation time of $T_1=315\pm16~$\textmu s for our shallow NV -- a timescale which is known to limit the NV's nuclear spin decoherence time\,\cite{Metsch2019a, Broadway2018a}.

\textbf{Scaling to NV ensembles}
An interesting question with particular relevance for potential applications in quantum sensing is whether our all-optical scheme also scales to ensembles of NV centers.
To address this question, we repeated our experiments on an ensemble of NV centers in an unstructured, CVD grown diamond sample with a preferential orientation of NV centers along one of the four possible crystal directions\,\cite{Lesik2015a} (see methods). 
We maximize the contrast $C$ in the same fashion as done in the single NV case, and determine $T_2^*$ through the full FID trace shown in Fig.\,\ref{Fig4}b.  
Using a least-squares fit as before, we find $C=2.08\pm0.04$\%, and $T_2^*=508.5\pm 17.4~$\textmu s for this NV ensemble.
While $C$ is comparable to the single NV case (with a slight deterioration due to the minority of non-aligned NVs), $T_2^*$ almost doubles.  
This value of $T_2^*$, however, is still short of the best reported values of $\approx2.2~$ms for $^{15}$NV nuclear spin coherence times\,\cite{Sangtawesin2016a, Jarmola2021a}. 
We exclude electron spin $T_1$ relaxation as a source for this fast nuclear spin decoherence, as we measured $T_1=5.8\pm0.5~$ms in this sample. 
Possible other sources for nuclear spin dephasing in our ensemble experiment include fluctuations in external magnetic field or temperature\,\cite{Broadway2018a, Sangtawesin2016a}.

     \section{Discussion}

Our all-optical $^{15}$N FID scheme lends itself to applications in quantum sensing, e.g. in magnetometry and gyroscopy (rotational sensing).
In the following we discuss the predicted performance of such all-optical coherent quantum sensing schemes.

The shot noise limited FID sensitivity for spin based, low-frequency magnetometry is given by\,\cite{Budker2007a, Degen2017a, Barry2020a} 
\begin{equation}
\label{sensitivity_mag}
\eta_{\rm mag} \approx \frac{1}{\gamma C \sqrt{N T_2^*}}\,.
\end{equation}
Here, $N$ is the average number of detected photons per readout pulse, $C$ is the readout contrast, and $\gamma$ is the gyromagnetic ration of the spins employed for sensing.  
Further sensitivity reductions due to overhead in preparation and measurement of quantum states are not included in this expression, but of little relevance to our conclusion, given the long $T_2^*$ times at hand.

Evaluating Eq.\,\eqref{sensitivity_mag} for our single NV data ($T_2^* = 250~$\textmu s, $C=4\%$, $N=0.1$) using the effective nuclear gyromagnetic ratio $\gamma = 1.2\gamma_I$ determined from the slope of the data in Fig.\,\ref{Fig1}e (or from Eq.\,\eqref{Beff}), we obtain a photon shot noise limited magnetometry sensitivity of $\eta_{\rm mag} = 154~\text{\textmu T}/\sqrt{\text{Hz}}$.
Further, given that our approach scales to NV ensembles, we make predictions on future ensemble NV magnetometry sensitivity. 
For this, we assume a laser power of $100~$mW, a $350~$ns readout window, and a conversion ratio of excitation photons to detected PL photons of $3.4~$\%\,\cite{Chatzidrosos2021a}, to obtain $N = 3.2\cdot10^9$. 
Together with the measured ensemble values $T_2^* = 500~$\textmu s and $C=2~\%$, we obtain $\eta_{\rm mag}^{\rm ensemble} = 1.22~$nT$/\sqrt{\rm Hz}$.

For spin based gyroscopy, the sensitivity is determined in analogy to Eq.\,\eqref{sensitivity_mag}, but with omission of the gyromagnetic ratio, i.e. $\eta_{\rm gyro}=\gamma\cdot\eta_{\rm mag}$\,\cite{Jarmola2020a}.
Nuclear spins are therefore particularly attractive for gyroscopy, since their long $T_2^*$ times generally offer them better sensitivities compared to electron spins, while they are less susceptible to magnetic fields and their fluctuations.
Employing the same procedure as before, we obtain a projected ensemble gyroscope sensitivity of $\eta_{\rm gyro}^{\rm ensemble} \approx 135~^\circ/\sqrt{\rm hour}$.

To place these estimates in context, we note that best reported magnetometry sensitivities using electron spin ensembles in diamond were $\tilde{\eta}_{\rm mag}^{\rm ensemble}<10$\,pT$/\sqrt{\rm Hz}$\,\cite{Zhang2021a}, while NV-based nuclear spin gyroscopes have recently achieved sensitivities $\tilde{\eta}_{\rm gyro}^{\rm ensemble}=280~^\circ/\sqrt{\rm hour}$\,\cite{Jarmola2021a}.
While for magnetometry, our projected sensitivity alone is not competitive with the state-of-the-art, the microwave-free modality we present still lends itself to specific applications, e.g. remote sensing through optical fibres\,\cite{Patel2020a}, or for cases where the MW drive would critically affect the sample of interest.
Conversely, for gyroscopy, we project numbers competitive with previous approaches. 
The added feature of all-optical NV gyroscopy is hereby a key asset, which may enable future integrated and power-efficient NV gyroscopes.

Looking forward, we note that our all-optical nuclear spin sensing scheme is also amenable to alternate high fidelity readout schemes to increase measurement contrast, based on spin-to-charge conversion\,\cite{Shields2015a, Hopper2018a}.
Another potential path to improving contrast $C$ is to dynamically pulse the field misalignment angle between spin initialisation and readout, to separately optimise the two processes.

In conclusion, we have presented an all-optical scheme for observing FID dynamics of $^{15}$N nuclear spins in diamond NV centers.  
Our technique is based on optical pumping of the $^{15}$N nuclear spin into a quantum superposition state -- a novel optical pumping process that occurs near the NV's ESLAC in presence of a small transverse magnetic field.   
These results may find applications in various fields of quantum sensing, most notably all-optical magnetometry and gyroscopy, for which we give benchmark comparisons which compare favourably with the state-of-the-art. 
Our results also suggest possible extensions to a range of other, relevant scenarios, including analogous dynamics near the NV's ground-state level anti-crossing, or all-optical addressing of nearby $^{13}C$ nuclear spins.
The nuclear spin dynamics we discussed should generally be observable in any color center exhibiting suitable level anti-crossing dynamics and coupling to nuclear spins, and might as such offer interesting opportunities for sensing with and characterization of novel color centers in a variety of solid state hosts.

	\section{Methods}

\textbf{Single NV diamond sample} The majority of our experimental results (Figs.\,\ref{Fig1}-\ref{Fig3} and Fig\,\ref{Fig4}a) were obtained on a single NV center that was created in an ``electronic grade'' diamond sample (Element Six) by ion implantation and subsequent sample annealing\,\cite{Chu2014a}.
For implantation, we employed singly charged $^{15}$N ions at a flux of $10^{11}~$cm$^{-2}$ and an energy of $6~$keV, corresponding to a nominal implantation depth of about $9~$nm\,\cite{Pezzagna2010a}.  
To increase PL collection efficiency, parabolic diamond pillars were fabricated in the diamond surface\,\cite{Hedrich2020a} subsequent to NV creation -- a single pillar containing an individual NV center was studied in this work.

\textbf{NV ensemble diamond sample} The NV ensemble sample used to obtain data shown in Fig.\,\ref{Fig4}b was grown on a CVD diamond substrate along the $(113)$ crystal orientation to facilitate Nitrogen incorporation and create NVs preferentially oriented along the NV-axis lying closest to the growth plane\,\cite{Lesik2015a}.
A $15$~\textmu m thick layer containing NVs was obtained using $^{12}$C and $^{15}$N enriched gas mixture, which led to an estimated NV density of $\sim300~$ppb\,\cite{Balasubramanian2022a}.

\textbf{Experimental setup} A home-built confocal microscope (Olympus LMPLFLN-$100$ objective, $\text{NA}=0.8$) was used to focus a green laser (Cobolt 06-MLD; emission wavelength $515~$nm) on the sample and to simultaneously collect the emitted red PL. 
All data shown in this paper were taken by optically exciting the NV(s) near saturation, which for our setup corresponded to a laser power of about $50~$\textmu W for the single NV in a nano-pillar, and $2.2~$mW for the ensemble of NVs in unstructured diamond.
A static magnetic field was applied using a permanent neodymium disk magnet (supermagnete, S-60-05-N) mounted on a linear translation stage, to tune the magnetic field strength at the NV location.   
For precise magnetic field alignment near the ESLAC, the magnet is mounted on a goniometric stage (SmarAct SGO-60.5 and SGO-77.5).  
Finally, the laser and photon-detectors were gated with pulses which were created and synchronized using a high-frequency signal generator (Zurich Instruments SHFSG), which also served as a source for microwave pulses used for the characterization of the magnetic field via optically detected magnetic resonance experiment.

	\section*{Acknowledgments}

We gratefully acknowledge Vincent Jacques, David Broadway, Andrey Jarmola, Sean Lourette and Dima Budker for fruitful discussions.  
We further acknowledge financial support through the NCCR QSIT (Grant No.$185902$), the Swiss Nanoscience Institute, and through the Swiss NSF Project Grant No.$188521$.

     \bibliographystyle{apsrev4-2}
		 \bibliography{AllOpticalNuclearQuantumSensing}

\begin{thebibliography}{56}%
\makeatletter
\providecommand \@ifxundefined [1]{%
 \@ifx{#1\undefined}
}%
\providecommand \@ifnum [1]{%
 \ifnum #1\expandafter \@firstoftwo
 \else \expandafter \@secondoftwo
 \fi
}%
\providecommand \@ifx [1]{%
 \ifx #1\expandafter \@firstoftwo
 \else \expandafter \@secondoftwo
 \fi
}%
\providecommand \natexlab [1]{#1}%
\providecommand \enquote  [1]{``#1''}%
\providecommand \bibnamefont  [1]{#1}%
\providecommand \bibfnamefont [1]{#1}%
\providecommand \citenamefont [1]{#1}%
\providecommand \href@noop [0]{\@secondoftwo}%
\providecommand \href [0]{\begingroup \@sanitize@url \@href}%
\providecommand \@href[1]{\@@startlink{#1}\@@href}%
\providecommand \@@href[1]{\endgroup#1\@@endlink}%
\providecommand \@sanitize@url [0]{\catcode `\\12\catcode `\$12\catcode
  `\&12\catcode `\#12\catcode `\^12\catcode `\_12\catcode `\%12\relax}%
\providecommand \@@startlink[1]{}%
\providecommand \@@endlink[0]{}%
\providecommand \url  [0]{\begingroup\@sanitize@url \@url }%
\providecommand \@url [1]{\endgroup\@href {#1}{\urlprefix }}%
\providecommand \urlprefix  [0]{URL }%
\providecommand \Eprint [0]{\href }%
\providecommand \doibase [0]{https://doi.org/}%
\providecommand \selectlanguage [0]{\@gobble}%
\providecommand \bibinfo  [0]{\@secondoftwo}%
\providecommand \bibfield  [0]{\@secondoftwo}%
\providecommand \translation [1]{[#1]}%
\providecommand \BibitemOpen [0]{}%
\providecommand \bibitemStop [0]{}%
\providecommand \bibitemNoStop [0]{.\EOS\space}%
\providecommand \EOS [0]{\spacefactor3000\relax}%
\providecommand \BibitemShut  [1]{\csname bibitem#1\endcsname}%
\let\auto@bib@innerbib\@empty
\bibitem [{\citenamefont {Taylor}\ \emph {et~al.}(2008)\citenamefont {Taylor},
  \citenamefont {Cappellaro}, \citenamefont {Childress}, \citenamefont {Jiang},
  \citenamefont {Budker}, \citenamefont {Hemmer}, \citenamefont {Yacoby},
  \citenamefont {Walsworth},\ and\ \citenamefont {Lukin}}]{Taylor2008a}%
  \BibitemOpen
  \bibfield  {author} {\bibinfo {author} {\bibfnamefont {J.}~\bibnamefont
  {Taylor}}, \bibinfo {author} {\bibfnamefont {P.}~\bibnamefont {Cappellaro}},
  \bibinfo {author} {\bibfnamefont {L.}~\bibnamefont {Childress}}, \bibinfo
  {author} {\bibfnamefont {L.}~\bibnamefont {Jiang}}, \bibinfo {author}
  {\bibfnamefont {D.}~\bibnamefont {Budker}}, \bibinfo {author} {\bibfnamefont
  {P.}~\bibnamefont {Hemmer}}, \bibinfo {author} {\bibfnamefont
  {A.}~\bibnamefont {Yacoby}}, \bibinfo {author} {\bibfnamefont
  {R.}~\bibnamefont {Walsworth}},\ and\ \bibinfo {author} {\bibfnamefont
  {M.}~\bibnamefont {Lukin}},\ }\href {http://dx.doi.org/10.1038/nphys1075}
  {\bibfield  {journal} {\bibinfo  {journal} {Nature Physics}\ }\textbf
  {\bibinfo {volume} {4}},\ \bibinfo {pages} {810} (\bibinfo {year}
  {2008})}\BibitemShut {NoStop}%
\bibitem [{\citenamefont {Dolde}\ \emph {et~al.}(2011)\citenamefont {Dolde},
  \citenamefont {Fedder}, \citenamefont {Doherty}, \citenamefont {N{\"o}bauer},
  \citenamefont {Rempp}, \citenamefont {Balasubramanian}, \citenamefont {Wolf},
  \citenamefont {Reinhard}, \citenamefont {Hollenberg}, \citenamefont
  {Jelezko},\ and\ \citenamefont {Wrachtrup}}]{Dolde2011a}%
  \BibitemOpen
  \bibfield  {author} {\bibinfo {author} {\bibfnamefont {F.}~\bibnamefont
  {Dolde}}, \bibinfo {author} {\bibfnamefont {H.}~\bibnamefont {Fedder}},
  \bibinfo {author} {\bibfnamefont {M.~W.}\ \bibnamefont {Doherty}}, \bibinfo
  {author} {\bibfnamefont {T.}~\bibnamefont {N{\"o}bauer}}, \bibinfo {author}
  {\bibfnamefont {F.}~\bibnamefont {Rempp}}, \bibinfo {author} {\bibfnamefont
  {G.}~\bibnamefont {Balasubramanian}}, \bibinfo {author} {\bibfnamefont
  {T.}~\bibnamefont {Wolf}}, \bibinfo {author} {\bibfnamefont {F.}~\bibnamefont
  {Reinhard}}, \bibinfo {author} {\bibfnamefont {L.~C.~L.}\ \bibnamefont
  {Hollenberg}}, \bibinfo {author} {\bibfnamefont {F.}~\bibnamefont
  {Jelezko}},\ and\ \bibinfo {author} {\bibfnamefont {J.}~\bibnamefont
  {Wrachtrup}},\ }\href {https://doi.org/10.1038/nphys1969} {\bibfield
  {journal} {\bibinfo  {journal} {Nature Physics}\ }\textbf {\bibinfo {volume}
  {7}},\ \bibinfo {pages} {459} (\bibinfo {year} {2011})}\BibitemShut {NoStop}%
\bibitem [{\citenamefont {Acosta}\ \emph {et~al.}(2010)\citenamefont {Acosta},
  \citenamefont {Bauch}, \citenamefont {Ledbetter}, \citenamefont {Waxman},
  \citenamefont {Bouchard},\ and\ \citenamefont {Budker}}]{Acosta2010a}%
  \BibitemOpen
  \bibfield  {author} {\bibinfo {author} {\bibfnamefont {V.~M.}\ \bibnamefont
  {Acosta}}, \bibinfo {author} {\bibfnamefont {E.}~\bibnamefont {Bauch}},
  \bibinfo {author} {\bibfnamefont {M.~P.}\ \bibnamefont {Ledbetter}}, \bibinfo
  {author} {\bibfnamefont {A.}~\bibnamefont {Waxman}}, \bibinfo {author}
  {\bibfnamefont {L.-S.}\ \bibnamefont {Bouchard}},\ and\ \bibinfo {author}
  {\bibfnamefont {D.}~\bibnamefont {Budker}},\ }\href
  {https://dx.doi.org/10.1103/PhysRevLett.104.070801} {\bibfield  {journal}
  {\bibinfo  {journal} {Physical Review Letters}\ }\textbf {\bibinfo {volume}
  {104}},\ \bibinfo {pages} {070801} (\bibinfo {year} {2010})}\BibitemShut
  {NoStop}%
\bibitem [{\citenamefont {Ledbetter}\ \emph {et~al.}(2012)\citenamefont
  {Ledbetter}, \citenamefont {Jensen}, \citenamefont {Fischer}, \citenamefont
  {Jarmola},\ and\ \citenamefont {Budker}}]{Ledbetter2012a}%
  \BibitemOpen
  \bibfield  {author} {\bibinfo {author} {\bibfnamefont {M.~P.}\ \bibnamefont
  {Ledbetter}}, \bibinfo {author} {\bibfnamefont {K.}~\bibnamefont {Jensen}},
  \bibinfo {author} {\bibfnamefont {R.}~\bibnamefont {Fischer}}, \bibinfo
  {author} {\bibfnamefont {A.}~\bibnamefont {Jarmola}},\ and\ \bibinfo {author}
  {\bibfnamefont {D.}~\bibnamefont {Budker}},\ }\href
  {https://doi.org/10.1103/PhysRevA.86.052116} {\bibfield  {journal} {\bibinfo
  {journal} {Physical Review A}\ }\textbf {\bibinfo {volume} {86}},\ \bibinfo
  {pages} {052116} (\bibinfo {year} {2012})}\BibitemShut {NoStop}%
\bibitem [{\citenamefont {Barry}\ \emph {et~al.}(2016)\citenamefont {Barry},
  \citenamefont {Turner}, \citenamefont {Schloss}, \citenamefont {Glenn},
  \citenamefont {Song}, \citenamefont {Lukin}, \citenamefont {Park},\ and\
  \citenamefont {Walsworth}}]{Barry2016a}%
  \BibitemOpen
  \bibfield  {author} {\bibinfo {author} {\bibfnamefont {J.~F.}\ \bibnamefont
  {Barry}}, \bibinfo {author} {\bibfnamefont {M.~J.}\ \bibnamefont {Turner}},
  \bibinfo {author} {\bibfnamefont {J.~M.}\ \bibnamefont {Schloss}}, \bibinfo
  {author} {\bibfnamefont {D.~R.}\ \bibnamefont {Glenn}}, \bibinfo {author}
  {\bibfnamefont {Y.}~\bibnamefont {Song}}, \bibinfo {author} {\bibfnamefont
  {M.~D.}\ \bibnamefont {Lukin}}, \bibinfo {author} {\bibfnamefont
  {H.}~\bibnamefont {Park}},\ and\ \bibinfo {author} {\bibfnamefont {R.~L.}\
  \bibnamefont {Walsworth}},\ }\href {https://doi.org/10.1073/pnas.1601513113}
  {\bibfield  {journal} {\bibinfo  {journal} {Proceedings of the National
  Academy of Sciences}\ }\textbf {\bibinfo {volume} {113}},\ \bibinfo {pages}
  {14133} (\bibinfo {year} {2016})}\BibitemShut {NoStop}%
\bibitem [{\citenamefont {Glenn}\ \emph {et~al.}(2017)\citenamefont {Glenn},
  \citenamefont {Fu}, \citenamefont {Kehayias}, \citenamefont {Le~Sage},
  \citenamefont {Lima}, \citenamefont {Weiss},\ and\ \citenamefont
  {Walsworth}}]{Glenn2017a}%
  \BibitemOpen
  \bibfield  {author} {\bibinfo {author} {\bibfnamefont {D.~R.}\ \bibnamefont
  {Glenn}}, \bibinfo {author} {\bibfnamefont {R.~R.}\ \bibnamefont {Fu}},
  \bibinfo {author} {\bibfnamefont {P.}~\bibnamefont {Kehayias}}, \bibinfo
  {author} {\bibfnamefont {D.}~\bibnamefont {Le~Sage}}, \bibinfo {author}
  {\bibfnamefont {E.~A.}\ \bibnamefont {Lima}}, \bibinfo {author}
  {\bibfnamefont {B.~P.}\ \bibnamefont {Weiss}},\ and\ \bibinfo {author}
  {\bibfnamefont {R.~L.}\ \bibnamefont {Walsworth}},\ }\href
  {https://doi.org/10.1002/2017GC006946} {\bibfield  {journal} {\bibinfo
  {journal} {Geochemistry, Geophysics, Geosystems}\ }\textbf {\bibinfo {volume}
  {18}},\ \bibinfo {pages} {3254} (\bibinfo {year} {2017})}\BibitemShut
  {NoStop}%
\bibitem [{\citenamefont {Levine}\ \emph {et~al.}(2019)\citenamefont {Levine},
  \citenamefont {Turner}, \citenamefont {Kehayias}, \citenamefont {Hart},
  \citenamefont {Langellier}, \citenamefont {Trubko}, \citenamefont {Glenn},
  \citenamefont {Fu},\ and\ \citenamefont {Walsworth}}]{Levine2019a}%
  \BibitemOpen
  \bibfield  {author} {\bibinfo {author} {\bibfnamefont {E.~V.}\ \bibnamefont
  {Levine}}, \bibinfo {author} {\bibfnamefont {M.~J.}\ \bibnamefont {Turner}},
  \bibinfo {author} {\bibfnamefont {P.}~\bibnamefont {Kehayias}}, \bibinfo
  {author} {\bibfnamefont {C.~A.}\ \bibnamefont {Hart}}, \bibinfo {author}
  {\bibfnamefont {N.}~\bibnamefont {Langellier}}, \bibinfo {author}
  {\bibfnamefont {R.}~\bibnamefont {Trubko}}, \bibinfo {author} {\bibfnamefont
  {D.~R.}\ \bibnamefont {Glenn}}, \bibinfo {author} {\bibfnamefont {R.~R.}\
  \bibnamefont {Fu}},\ and\ \bibinfo {author} {\bibfnamefont {R.~L.}\
  \bibnamefont {Walsworth}},\ }\href {https://doi.org/10.1515/nanoph-2019-0209}
  {\bibfield  {journal} {\bibinfo  {journal} {Nanophotonics}\ }\textbf
  {\bibinfo {volume} {8}},\ \bibinfo {pages} {1945} (\bibinfo {year}
  {2019})}\BibitemShut {NoStop}%
\bibitem [{\citenamefont {Phillips}\ \emph {et~al.}(2020)\citenamefont
  {Phillips}, \citenamefont {Wright}, \citenamefont {Kiss-Toth}, \citenamefont
  {Read}, \citenamefont {Riou}, \citenamefont {Maddox}, \citenamefont
  {Maskell},\ and\ \citenamefont {Ralph}}]{Phillips2020a}%
  \BibitemOpen
  \bibfield  {author} {\bibinfo {author} {\bibfnamefont {A.~M.}\ \bibnamefont
  {Phillips}}, \bibinfo {author} {\bibfnamefont {M.~J.}\ \bibnamefont
  {Wright}}, \bibinfo {author} {\bibfnamefont {M.}~\bibnamefont {Kiss-Toth}},
  \bibinfo {author} {\bibfnamefont {I.}~\bibnamefont {Read}}, \bibinfo {author}
  {\bibfnamefont {I.}~\bibnamefont {Riou}}, \bibinfo {author} {\bibfnamefont
  {S.}~\bibnamefont {Maddox}}, \bibinfo {author} {\bibfnamefont
  {S.}~\bibnamefont {Maskell}},\ and\ \bibinfo {author} {\bibfnamefont {J.~F.}\
  \bibnamefont {Ralph}},\ }in\ \href {https://doi.org/10.1117/12.2583999}
  {\emph {\bibinfo {booktitle} {Cold {Atoms} for {Quantum} {Technologies}}}},\
  Vol.\ \bibinfo {volume} {11578}\ (\bibinfo  {publisher} {SPIE},\ \bibinfo
  {year} {2020})\ p.\ \bibinfo {pages} {115780C}\BibitemShut {NoStop}%
\bibitem [{\citenamefont {Casola}\ \emph {et~al.}(2018)\citenamefont {Casola},
  \citenamefont {van~der Sar},\ and\ \citenamefont {Yacoby}}]{Casola2018a}%
  \BibitemOpen
  \bibfield  {author} {\bibinfo {author} {\bibfnamefont {F.}~\bibnamefont
  {Casola}}, \bibinfo {author} {\bibfnamefont {T.}~\bibnamefont {van~der
  Sar}},\ and\ \bibinfo {author} {\bibfnamefont {A.}~\bibnamefont {Yacoby}},\
  }\href {https://doi.org/10.1038/natrevmats.2017.88} {\bibfield  {journal}
  {\bibinfo  {journal} {Nature Reviews Materials}\ }\textbf {\bibinfo {volume}
  {3}},\ \bibinfo {pages} {1} (\bibinfo {year} {2018})}\BibitemShut {NoStop}%
\bibitem [{\citenamefont {Doherty}\ \emph {et~al.}(2013)\citenamefont
  {Doherty}, \citenamefont {Manson}, \citenamefont {Delaney}, \citenamefont
  {Jelezko}, \citenamefont {Wrachtrup},\ and\ \citenamefont
  {Hollenberg}}]{Doherty2013a}%
  \BibitemOpen
  \bibfield  {author} {\bibinfo {author} {\bibfnamefont {M.~W.}\ \bibnamefont
  {Doherty}}, \bibinfo {author} {\bibfnamefont {N.~B.}\ \bibnamefont {Manson}},
  \bibinfo {author} {\bibfnamefont {P.}~\bibnamefont {Delaney}}, \bibinfo
  {author} {\bibfnamefont {F.}~\bibnamefont {Jelezko}}, \bibinfo {author}
  {\bibfnamefont {J.}~\bibnamefont {Wrachtrup}},\ and\ \bibinfo {author}
  {\bibfnamefont {L.~C.~L.}\ \bibnamefont {Hollenberg}},\ }\href
  {https://doi.org/10.1016/j.physrep.2013.02.001} {\bibfield  {journal}
  {\bibinfo  {journal} {Physics Reports}\ }\bibinfo {series} {The
  nitrogen-vacancy colour centre in diamond},\ \textbf {\bibinfo {volume}
  {528}},\ \bibinfo {pages} {1} (\bibinfo {year} {2013})}\BibitemShut {NoStop}%
\bibitem [{\citenamefont {Balasubramanian}\ \emph {et~al.}(2009)\citenamefont
  {Balasubramanian}, \citenamefont {Neumann}, \citenamefont {Twitchen},
  \citenamefont {Markham}, \citenamefont {Kolesov}, \citenamefont {Mizuochi},
  \citenamefont {Isoya}, \citenamefont {Achard}, \citenamefont {Beck},
  \citenamefont {Tissler}, \citenamefont {Jacques}, \citenamefont {Hemmer},
  \citenamefont {Jelezko},\ and\ \citenamefont
  {Wrachtrup}}]{Balasubramanian2009a}%
  \BibitemOpen
  \bibfield  {author} {\bibinfo {author} {\bibfnamefont {G.}~\bibnamefont
  {Balasubramanian}}, \bibinfo {author} {\bibfnamefont {P.}~\bibnamefont
  {Neumann}}, \bibinfo {author} {\bibfnamefont {D.}~\bibnamefont {Twitchen}},
  \bibinfo {author} {\bibfnamefont {M.}~\bibnamefont {Markham}}, \bibinfo
  {author} {\bibfnamefont {R.}~\bibnamefont {Kolesov}}, \bibinfo {author}
  {\bibfnamefont {N.}~\bibnamefont {Mizuochi}}, \bibinfo {author}
  {\bibfnamefont {J.}~\bibnamefont {Isoya}}, \bibinfo {author} {\bibfnamefont
  {J.}~\bibnamefont {Achard}}, \bibinfo {author} {\bibfnamefont
  {J.}~\bibnamefont {Beck}}, \bibinfo {author} {\bibfnamefont {J.}~\bibnamefont
  {Tissler}}, \bibinfo {author} {\bibfnamefont {V.}~\bibnamefont {Jacques}},
  \bibinfo {author} {\bibfnamefont {P.~R.}\ \bibnamefont {Hemmer}}, \bibinfo
  {author} {\bibfnamefont {F.}~\bibnamefont {Jelezko}},\ and\ \bibinfo {author}
  {\bibfnamefont {J.}~\bibnamefont {Wrachtrup}},\ }\href
  {https://doi.org/10.1038/nmat2420} {\bibfield  {journal} {\bibinfo  {journal}
  {Nature Materials}\ }\textbf {\bibinfo {volume} {8}},\ \bibinfo {pages} {383}
  (\bibinfo {year} {2009})}\BibitemShut {NoStop}%
\bibitem [{\citenamefont {Bar-Gill}\ \emph {et~al.}(2013)\citenamefont
  {Bar-Gill}, \citenamefont {Pham}, \citenamefont {Jarmola}, \citenamefont
  {Budker},\ and\ \citenamefont {Walsworth}}]{BarGill2013a}%
  \BibitemOpen
  \bibfield  {author} {\bibinfo {author} {\bibfnamefont {N.}~\bibnamefont
  {Bar-Gill}}, \bibinfo {author} {\bibfnamefont {L.~M.}\ \bibnamefont {Pham}},
  \bibinfo {author} {\bibfnamefont {A.}~\bibnamefont {Jarmola}}, \bibinfo
  {author} {\bibfnamefont {D.}~\bibnamefont {Budker}},\ and\ \bibinfo {author}
  {\bibfnamefont {R.~L.}\ \bibnamefont {Walsworth}},\ }\href
  {https://doi.org/10.1038/ncomms2771} {\bibfield  {journal} {\bibinfo
  {journal} {Nature Communications}\ }\textbf {\bibinfo {volume} {4}},\
  \bibinfo {pages} {1743} (\bibinfo {year} {2013})}\BibitemShut {NoStop}%
\bibitem [{\citenamefont {Kennedy}\ \emph {et~al.}(2003)\citenamefont
  {Kennedy}, \citenamefont {Colton}, \citenamefont {Butler}, \citenamefont
  {Linares},\ and\ \citenamefont {Doering}}]{Kennedy2003a}%
  \BibitemOpen
  \bibfield  {author} {\bibinfo {author} {\bibfnamefont {T.~A.}\ \bibnamefont
  {Kennedy}}, \bibinfo {author} {\bibfnamefont {J.~S.}\ \bibnamefont {Colton}},
  \bibinfo {author} {\bibfnamefont {J.~E.}\ \bibnamefont {Butler}}, \bibinfo
  {author} {\bibfnamefont {R.~C.}\ \bibnamefont {Linares}},\ and\ \bibinfo
  {author} {\bibfnamefont {P.~J.}\ \bibnamefont {Doering}},\ }\href
  {https://doi.org/10.1063/1.1626791} {\bibfield  {journal} {\bibinfo
  {journal} {Applied Physics Letters}\ }\textbf {\bibinfo {volume} {83}},\
  \bibinfo {pages} {4190} (\bibinfo {year} {2003})}\BibitemShut {NoStop}%
\bibitem [{\citenamefont {Gruber}\ \emph {et~al.}(1997)\citenamefont {Gruber},
  \citenamefont {Dr{\"a}benstedt}, \citenamefont {Tietz}, \citenamefont
  {Fleury}, \citenamefont {Wrachtrup},\ and\ \citenamefont
  {Borczyskowski}}]{Gruber1997a}%
  \BibitemOpen
  \bibfield  {author} {\bibinfo {author} {\bibfnamefont {A.}~\bibnamefont
  {Gruber}}, \bibinfo {author} {\bibfnamefont {A.}~\bibnamefont
  {Dr{\"a}benstedt}}, \bibinfo {author} {\bibfnamefont {C.}~\bibnamefont
  {Tietz}}, \bibinfo {author} {\bibfnamefont {L.}~\bibnamefont {Fleury}},
  \bibinfo {author} {\bibfnamefont {J.}~\bibnamefont {Wrachtrup}},\ and\
  \bibinfo {author} {\bibfnamefont {C.~v.}\ \bibnamefont {Borczyskowski}},\
  }\href {https://doi.org/10.1126/science.276.5321.2012} {\bibfield  {journal}
  {\bibinfo  {journal} {Science}\ }\textbf {\bibinfo {volume} {276}},\ \bibinfo
  {pages} {2012} (\bibinfo {year} {1997})}\BibitemShut {NoStop}%
\bibitem [{\citenamefont {Harrison}\ \emph {et~al.}(2004)\citenamefont
  {Harrison}, \citenamefont {Sellars},\ and\ \citenamefont
  {Manson}}]{Harrison2004a}%
  \BibitemOpen
  \bibfield  {author} {\bibinfo {author} {\bibfnamefont {J.}~\bibnamefont
  {Harrison}}, \bibinfo {author} {\bibfnamefont {M.~J.}\ \bibnamefont
  {Sellars}},\ and\ \bibinfo {author} {\bibfnamefont {N.~B.}\ \bibnamefont
  {Manson}},\ }\href {https://doi.org/10.1016/j.jlumin.2003.12.020} {\bibfield
  {journal} {\bibinfo  {journal} {Journal of Luminescence}\ }\bibinfo {series}
  {Proceedings of the 8th {International} {Meeting} on {Hole} {Burning},
  {Single} {Molecule}, and {Related} {Spectroscopies}: {Science} and
  {Applications}},\ \textbf {\bibinfo {volume} {107}},\ \bibinfo {pages} {245}
  (\bibinfo {year} {2004})}\BibitemShut {NoStop}%
\bibitem [{\citenamefont {Tetienne}\ \emph {et~al.}(2012)\citenamefont
  {Tetienne}, \citenamefont {Rondin}, \citenamefont {Spinicelli}, \citenamefont
  {Chipaux}, \citenamefont {Debuisschert}, \citenamefont {Roch},\ and\
  \citenamefont {Jacques}}]{Tetienne2012a}%
  \BibitemOpen
  \bibfield  {author} {\bibinfo {author} {\bibfnamefont {J.-P.}\ \bibnamefont
  {Tetienne}}, \bibinfo {author} {\bibfnamefont {L.}~\bibnamefont {Rondin}},
  \bibinfo {author} {\bibfnamefont {P.}~\bibnamefont {Spinicelli}}, \bibinfo
  {author} {\bibfnamefont {M.}~\bibnamefont {Chipaux}}, \bibinfo {author}
  {\bibfnamefont {T.}~\bibnamefont {Debuisschert}}, \bibinfo {author}
  {\bibfnamefont {J.-F.}\ \bibnamefont {Roch}},\ and\ \bibinfo {author}
  {\bibfnamefont {V.}~\bibnamefont {Jacques}},\ }\href
  {https://doi.org/10.1088/1367-2630/14/10/103033} {\bibfield  {journal}
  {\bibinfo  {journal} {New Journal of Physics}\ }\textbf {\bibinfo {volume}
  {14}},\ \bibinfo {pages} {103033} (\bibinfo {year} {2012})}\BibitemShut
  {NoStop}%
\bibitem [{\citenamefont {Degen}\ \emph {et~al.}(2017)\citenamefont {Degen},
  \citenamefont {Reinhard},\ and\ \citenamefont {Cappellaro}}]{Degen2017a}%
  \BibitemOpen
  \bibfield  {author} {\bibinfo {author} {\bibfnamefont {C.}~\bibnamefont
  {Degen}}, \bibinfo {author} {\bibfnamefont {F.}~\bibnamefont {Reinhard}},\
  and\ \bibinfo {author} {\bibfnamefont {P.}~\bibnamefont {Cappellaro}},\
  }\href {https://doi.org/10.1103/RevModPhys.89.035002} {\bibfield  {journal}
  {\bibinfo  {journal} {Reviews of Modern Physics}\ }\textbf {\bibinfo {volume}
  {89}},\ \bibinfo {pages} {035002} (\bibinfo {year} {2017})}\BibitemShut
  {NoStop}%
\bibitem [{\citenamefont {Waldherr}\ \emph {et~al.}(2012)\citenamefont
  {Waldherr}, \citenamefont {Beck}, \citenamefont {Neumann}, \citenamefont
  {Said}, \citenamefont {Nitsche}, \citenamefont {Markham}, \citenamefont
  {Twitchen}, \citenamefont {Twamley}, \citenamefont {Jelezko},\ and\
  \citenamefont {Wrachtrup}}]{Waldherr2012a}%
  \BibitemOpen
  \bibfield  {author} {\bibinfo {author} {\bibfnamefont {G.}~\bibnamefont
  {Waldherr}}, \bibinfo {author} {\bibfnamefont {J.}~\bibnamefont {Beck}},
  \bibinfo {author} {\bibfnamefont {P.}~\bibnamefont {Neumann}}, \bibinfo
  {author} {\bibfnamefont {R.~S.}\ \bibnamefont {Said}}, \bibinfo {author}
  {\bibfnamefont {M.}~\bibnamefont {Nitsche}}, \bibinfo {author} {\bibfnamefont
  {M.~L.}\ \bibnamefont {Markham}}, \bibinfo {author} {\bibfnamefont {D.~J.}\
  \bibnamefont {Twitchen}}, \bibinfo {author} {\bibfnamefont {J.}~\bibnamefont
  {Twamley}}, \bibinfo {author} {\bibfnamefont {F.}~\bibnamefont {Jelezko}},\
  and\ \bibinfo {author} {\bibfnamefont {J.}~\bibnamefont {Wrachtrup}},\ }\href
  {https://doi.org/10.1038/nnano.2011.224} {\bibfield  {journal} {\bibinfo
  {journal} {Nature Nanotechnology}\ }\textbf {\bibinfo {volume} {7}},\
  \bibinfo {pages} {105} (\bibinfo {year} {2012})}\BibitemShut {NoStop}%
\bibitem [{\citenamefont {Pfender}\ \emph {et~al.}(2017)\citenamefont
  {Pfender}, \citenamefont {Aslam}, \citenamefont {Sumiya}, \citenamefont
  {Onoda}, \citenamefont {Neumann}, \citenamefont {Isoya}, \citenamefont
  {Meriles},\ and\ \citenamefont {Wrachtrup}}]{Pfender2017a}%
  \BibitemOpen
  \bibfield  {author} {\bibinfo {author} {\bibfnamefont {M.}~\bibnamefont
  {Pfender}}, \bibinfo {author} {\bibfnamefont {N.}~\bibnamefont {Aslam}},
  \bibinfo {author} {\bibfnamefont {H.}~\bibnamefont {Sumiya}}, \bibinfo
  {author} {\bibfnamefont {S.}~\bibnamefont {Onoda}}, \bibinfo {author}
  {\bibfnamefont {P.}~\bibnamefont {Neumann}}, \bibinfo {author} {\bibfnamefont
  {J.}~\bibnamefont {Isoya}}, \bibinfo {author} {\bibfnamefont {C.~A.}\
  \bibnamefont {Meriles}},\ and\ \bibinfo {author} {\bibfnamefont
  {J.}~\bibnamefont {Wrachtrup}},\ }\href
  {https://doi.org/10.1038/s41467-017-00964-z} {\bibfield  {journal} {\bibinfo
  {journal} {Nature Communications}\ }\textbf {\bibinfo {volume} {8}},\
  \bibinfo {pages} {834} (\bibinfo {year} {2017})}\BibitemShut {NoStop}%
\bibitem [{\citenamefont {Pompili}\ \emph {et~al.}(2021)\citenamefont
  {Pompili}, \citenamefont {Hermans}, \citenamefont {Baier}, \citenamefont
  {Beukers}, \citenamefont {Humphreys}, \citenamefont {Schouten}, \citenamefont
  {Vermeulen}, \citenamefont {Tiggelman}, \citenamefont {dos Santos~Martins},
  \citenamefont {Dirkse}, \citenamefont {Wehner},\ and\ \citenamefont
  {Hanson}}]{Pompili2021a}%
  \BibitemOpen
  \bibfield  {author} {\bibinfo {author} {\bibfnamefont {M.}~\bibnamefont
  {Pompili}}, \bibinfo {author} {\bibfnamefont {S.~L.~N.}\ \bibnamefont
  {Hermans}}, \bibinfo {author} {\bibfnamefont {S.}~\bibnamefont {Baier}},
  \bibinfo {author} {\bibfnamefont {H.~K.~C.}\ \bibnamefont {Beukers}},
  \bibinfo {author} {\bibfnamefont {P.~C.}\ \bibnamefont {Humphreys}}, \bibinfo
  {author} {\bibfnamefont {R.~N.}\ \bibnamefont {Schouten}}, \bibinfo {author}
  {\bibfnamefont {R.~F.~L.}\ \bibnamefont {Vermeulen}}, \bibinfo {author}
  {\bibfnamefont {M.~J.}\ \bibnamefont {Tiggelman}}, \bibinfo {author}
  {\bibfnamefont {L.}~\bibnamefont {dos Santos~Martins}}, \bibinfo {author}
  {\bibfnamefont {B.}~\bibnamefont {Dirkse}}, \bibinfo {author} {\bibfnamefont
  {S.}~\bibnamefont {Wehner}},\ and\ \bibinfo {author} {\bibfnamefont
  {R.}~\bibnamefont {Hanson}},\ }\href
  {https://doi.org/10.1126/science.abg1919} {\bibfield  {journal} {\bibinfo
  {journal} {Science}\ }\textbf {\bibinfo {volume} {372}},\ \bibinfo {pages}
  {259} (\bibinfo {year} {2021})}\BibitemShut {NoStop}%
\bibitem [{\citenamefont {Lovchinsky}\ \emph {et~al.}(2016)\citenamefont
  {Lovchinsky}, \citenamefont {Sushkov}, \citenamefont {Urbach}, \citenamefont
  {de~Leon}, \citenamefont {Choi}, \citenamefont {De~Greve}, \citenamefont
  {Evans}, \citenamefont {Gertner}, \citenamefont {Bersin}, \citenamefont
  {M{\"u}ller}, \citenamefont {McGuinness}, \citenamefont {Jelezko},
  \citenamefont {Walsworth}, \citenamefont {Park},\ and\ \citenamefont
  {Lukin}}]{Lovchinsky2016a}%
  \BibitemOpen
  \bibfield  {author} {\bibinfo {author} {\bibfnamefont {I.}~\bibnamefont
  {Lovchinsky}}, \bibinfo {author} {\bibfnamefont {A.~O.}\ \bibnamefont
  {Sushkov}}, \bibinfo {author} {\bibfnamefont {E.}~\bibnamefont {Urbach}},
  \bibinfo {author} {\bibfnamefont {N.~P.}\ \bibnamefont {de~Leon}}, \bibinfo
  {author} {\bibfnamefont {S.}~\bibnamefont {Choi}}, \bibinfo {author}
  {\bibfnamefont {K.}~\bibnamefont {De~Greve}}, \bibinfo {author}
  {\bibfnamefont {R.}~\bibnamefont {Evans}}, \bibinfo {author} {\bibfnamefont
  {R.}~\bibnamefont {Gertner}}, \bibinfo {author} {\bibfnamefont
  {E.}~\bibnamefont {Bersin}}, \bibinfo {author} {\bibfnamefont
  {C.}~\bibnamefont {M{\"u}ller}}, \bibinfo {author} {\bibfnamefont
  {L.}~\bibnamefont {McGuinness}}, \bibinfo {author} {\bibfnamefont
  {F.}~\bibnamefont {Jelezko}}, \bibinfo {author} {\bibfnamefont {R.~L.}\
  \bibnamefont {Walsworth}}, \bibinfo {author} {\bibfnamefont {H.}~\bibnamefont
  {Park}},\ and\ \bibinfo {author} {\bibfnamefont {M.~D.}\ \bibnamefont
  {Lukin}},\ }\href {https://doi.org/10.1126/science.aad8022} {\bibfield
  {journal} {\bibinfo  {journal} {Science}\ }\textbf {\bibinfo {volume}
  {351}},\ \bibinfo {pages} {836} (\bibinfo {year} {2016})}\BibitemShut
  {NoStop}%
\bibitem [{\citenamefont {Chen}\ \emph {et~al.}(2013)\citenamefont {Chen},
  \citenamefont {Sun}, \citenamefont {Zou}, \citenamefont {Cui}, \citenamefont
  {Zhou},\ and\ \citenamefont {Guo}}]{Chen2013a}%
  \BibitemOpen
  \bibfield  {author} {\bibinfo {author} {\bibfnamefont {X.-D.}\ \bibnamefont
  {Chen}}, \bibinfo {author} {\bibfnamefont {F.-W.}\ \bibnamefont {Sun}},
  \bibinfo {author} {\bibfnamefont {C.-L.}\ \bibnamefont {Zou}}, \bibinfo
  {author} {\bibfnamefont {J.-M.}\ \bibnamefont {Cui}}, \bibinfo {author}
  {\bibfnamefont {L.-M.}\ \bibnamefont {Zhou}},\ and\ \bibinfo {author}
  {\bibfnamefont {G.-C.}\ \bibnamefont {Guo}},\ }\href
  {https://doi.org/10.1209/0295-5075/101/67003} {\bibfield  {journal} {\bibinfo
   {journal} {EPL (Europhysics Letters)}\ }\textbf {\bibinfo {volume} {101}},\
  \bibinfo {pages} {67003} (\bibinfo {year} {2013})}\BibitemShut {NoStop}%
\bibitem [{\citenamefont {Jakobi}\ \emph {et~al.}(2017)\citenamefont {Jakobi},
  \citenamefont {Neumann}, \citenamefont {Wang}, \citenamefont {Dasari},
  \citenamefont {El~Hallak}, \citenamefont {Bashir}, \citenamefont {Markham},
  \citenamefont {Edmonds}, \citenamefont {Twitchen},\ and\ \citenamefont
  {Wrachtrup}}]{Jakobi2017a}%
  \BibitemOpen
  \bibfield  {author} {\bibinfo {author} {\bibfnamefont {I.}~\bibnamefont
  {Jakobi}}, \bibinfo {author} {\bibfnamefont {P.}~\bibnamefont {Neumann}},
  \bibinfo {author} {\bibfnamefont {Y.}~\bibnamefont {Wang}}, \bibinfo {author}
  {\bibfnamefont {D.~B.~R.}\ \bibnamefont {Dasari}}, \bibinfo {author}
  {\bibfnamefont {F.}~\bibnamefont {El~Hallak}}, \bibinfo {author}
  {\bibfnamefont {M.~A.}\ \bibnamefont {Bashir}}, \bibinfo {author}
  {\bibfnamefont {M.}~\bibnamefont {Markham}}, \bibinfo {author} {\bibfnamefont
  {A.}~\bibnamefont {Edmonds}}, \bibinfo {author} {\bibfnamefont
  {D.}~\bibnamefont {Twitchen}},\ and\ \bibinfo {author} {\bibfnamefont
  {J.}~\bibnamefont {Wrachtrup}},\ }\href
  {https://doi.org/10.1038/nnano.2016.163} {\bibfield  {journal} {\bibinfo
  {journal} {Nature Nanotechnology}\ }\textbf {\bibinfo {volume} {12}},\
  \bibinfo {pages} {67} (\bibinfo {year} {2017})}\BibitemShut {NoStop}%
\bibitem [{\citenamefont {Sahin}\ \emph {et~al.}(2022)\citenamefont {Sahin},
  \citenamefont {de~Leon~Sanchez}, \citenamefont {Conti}, \citenamefont
  {Akkiraju}, \citenamefont {Reshetikhin}, \citenamefont {Druga}, \citenamefont
  {Aggarwal}, \citenamefont {Gilbert}, \citenamefont {Bhave},\ and\
  \citenamefont {Ajoy}}]{Sahin2022a}%
  \BibitemOpen
  \bibfield  {author} {\bibinfo {author} {\bibfnamefont {O.}~\bibnamefont
  {Sahin}}, \bibinfo {author} {\bibfnamefont {E.}~\bibnamefont
  {de~Leon~Sanchez}}, \bibinfo {author} {\bibfnamefont {S.}~\bibnamefont
  {Conti}}, \bibinfo {author} {\bibfnamefont {A.}~\bibnamefont {Akkiraju}},
  \bibinfo {author} {\bibfnamefont {P.}~\bibnamefont {Reshetikhin}}, \bibinfo
  {author} {\bibfnamefont {E.}~\bibnamefont {Druga}}, \bibinfo {author}
  {\bibfnamefont {A.}~\bibnamefont {Aggarwal}}, \bibinfo {author}
  {\bibfnamefont {B.}~\bibnamefont {Gilbert}}, \bibinfo {author} {\bibfnamefont
  {S.}~\bibnamefont {Bhave}},\ and\ \bibinfo {author} {\bibfnamefont
  {A.}~\bibnamefont {Ajoy}},\ }\href
  {https://doi.org/10.1038/s41467-022-32907-8} {\bibfield  {journal} {\bibinfo
  {journal} {Nature Communications}\ }\textbf {\bibinfo {volume} {13}},\
  \bibinfo {pages} {5486} (\bibinfo {year} {2022})}\BibitemShut {NoStop}%
\bibitem [{\citenamefont {Ajoy}\ and\ \citenamefont
  {Cappellaro}(2012)}]{Ajoy2012a}%
  \BibitemOpen
  \bibfield  {author} {\bibinfo {author} {\bibfnamefont {A.}~\bibnamefont
  {Ajoy}}\ and\ \bibinfo {author} {\bibfnamefont {P.}~\bibnamefont
  {Cappellaro}},\ }\href {https://doi.org/10.1103/PhysRevA.86.062104}
  {\bibfield  {journal} {\bibinfo  {journal} {Physical Review A}\ }\textbf
  {\bibinfo {volume} {86}},\ \bibinfo {pages} {062104} (\bibinfo {year}
  {2012})}\BibitemShut {NoStop}%
\bibitem [{\citenamefont {Jarmola}\ \emph {et~al.}(2021)\citenamefont
  {Jarmola}, \citenamefont {Lourette}, \citenamefont {Acosta}, \citenamefont
  {Birdwell}, \citenamefont {Bl{\"u}mler}, \citenamefont {Budker},
  \citenamefont {Ivanov},\ and\ \citenamefont {Malinovsky}}]{Jarmola2021a}%
  \BibitemOpen
  \bibfield  {author} {\bibinfo {author} {\bibfnamefont {A.}~\bibnamefont
  {Jarmola}}, \bibinfo {author} {\bibfnamefont {S.}~\bibnamefont {Lourette}},
  \bibinfo {author} {\bibfnamefont {V.~M.}\ \bibnamefont {Acosta}}, \bibinfo
  {author} {\bibfnamefont {A.~G.}\ \bibnamefont {Birdwell}}, \bibinfo {author}
  {\bibfnamefont {P.}~\bibnamefont {Bl{\"u}mler}}, \bibinfo {author}
  {\bibfnamefont {D.}~\bibnamefont {Budker}}, \bibinfo {author} {\bibfnamefont
  {T.}~\bibnamefont {Ivanov}},\ and\ \bibinfo {author} {\bibfnamefont {V.~S.}\
  \bibnamefont {Malinovsky}},\ }\href {https://doi.org/10.1126/sciadv.abl3840}
  {\bibfield  {journal} {\bibinfo  {journal} {Science Advances}\ }\textbf
  {\bibinfo {volume} {7}},\ \bibinfo {pages} {eabl3840} (\bibinfo {year}
  {2021})}\BibitemShut {NoStop}%
\bibitem [{\citenamefont {Wood}\ \emph {et~al.}(2016)\citenamefont {Wood},
  \citenamefont {Broadway}, \citenamefont {Hall}, \citenamefont {Stacey},
  \citenamefont {Simpson}, \citenamefont {Tetienne},\ and\ \citenamefont
  {Hollenberg}}]{Wood2016a}%
  \BibitemOpen
  \bibfield  {author} {\bibinfo {author} {\bibfnamefont {J.~D.~A.}\
  \bibnamefont {Wood}}, \bibinfo {author} {\bibfnamefont {D.~A.}\ \bibnamefont
  {Broadway}}, \bibinfo {author} {\bibfnamefont {L.~T.}\ \bibnamefont {Hall}},
  \bibinfo {author} {\bibfnamefont {A.}~\bibnamefont {Stacey}}, \bibinfo
  {author} {\bibfnamefont {D.~A.}\ \bibnamefont {Simpson}}, \bibinfo {author}
  {\bibfnamefont {J.-P.}\ \bibnamefont {Tetienne}},\ and\ \bibinfo {author}
  {\bibfnamefont {L.~C.~L.}\ \bibnamefont {Hollenberg}},\ }\href
  {https://doi.org/10.1103/PhysRevB.94.155402} {\bibfield  {journal} {\bibinfo
  {journal} {Phys. Rev. B}\ }\textbf {\bibinfo {volume} {94}},\ \bibinfo
  {pages} {155402} (\bibinfo {year} {2016})}\BibitemShut {NoStop}%
\bibitem [{\citenamefont {Jarmola}\ \emph {et~al.}(2020)\citenamefont
  {Jarmola}, \citenamefont {Fescenko}, \citenamefont {Acosta}, \citenamefont
  {Doherty}, \citenamefont {Fatemi}, \citenamefont {Ivanov}, \citenamefont
  {Budker},\ and\ \citenamefont {Malinovsky}}]{Jarmola2020a}%
  \BibitemOpen
  \bibfield  {author} {\bibinfo {author} {\bibfnamefont {A.}~\bibnamefont
  {Jarmola}}, \bibinfo {author} {\bibfnamefont {I.}~\bibnamefont {Fescenko}},
  \bibinfo {author} {\bibfnamefont {V.~M.}\ \bibnamefont {Acosta}}, \bibinfo
  {author} {\bibfnamefont {M.~W.}\ \bibnamefont {Doherty}}, \bibinfo {author}
  {\bibfnamefont {F.~K.}\ \bibnamefont {Fatemi}}, \bibinfo {author}
  {\bibfnamefont {T.}~\bibnamefont {Ivanov}}, \bibinfo {author} {\bibfnamefont
  {D.}~\bibnamefont {Budker}},\ and\ \bibinfo {author} {\bibfnamefont {V.~S.}\
  \bibnamefont {Malinovsky}},\ }\href
  {https://doi.org/10.1103/PhysRevResearch.2.023094} {\bibfield  {journal}
  {\bibinfo  {journal} {Physical Review Research}\ }\textbf {\bibinfo {volume}
  {2}},\ \bibinfo {pages} {023094} (\bibinfo {year} {2020})}\BibitemShut
  {NoStop}%
\bibitem [{\citenamefont {Jacques}\ \emph {et~al.}(2009)\citenamefont
  {Jacques}, \citenamefont {Neumann}, \citenamefont {Beck}, \citenamefont
  {Markham}, \citenamefont {Twitchen}, \citenamefont {Meijer}, \citenamefont
  {Kaiser}, \citenamefont {Balasubramanian}, \citenamefont {Jelezko},\ and\
  \citenamefont {Wrachtrup}}]{Jacques2009a}%
  \BibitemOpen
  \bibfield  {author} {\bibinfo {author} {\bibfnamefont {V.}~\bibnamefont
  {Jacques}}, \bibinfo {author} {\bibfnamefont {P.}~\bibnamefont {Neumann}},
  \bibinfo {author} {\bibfnamefont {J.}~\bibnamefont {Beck}}, \bibinfo {author}
  {\bibfnamefont {M.}~\bibnamefont {Markham}}, \bibinfo {author} {\bibfnamefont
  {D.}~\bibnamefont {Twitchen}}, \bibinfo {author} {\bibfnamefont
  {J.}~\bibnamefont {Meijer}}, \bibinfo {author} {\bibfnamefont
  {F.}~\bibnamefont {Kaiser}}, \bibinfo {author} {\bibfnamefont
  {G.}~\bibnamefont {Balasubramanian}}, \bibinfo {author} {\bibfnamefont
  {F.}~\bibnamefont {Jelezko}},\ and\ \bibinfo {author} {\bibfnamefont
  {J.}~\bibnamefont {Wrachtrup}},\ }\href
  {https://doi.org/10.1103/PhysRevLett.102.057403} {\bibfield  {journal}
  {\bibinfo  {journal} {Physical Review Letters}\ }\textbf {\bibinfo {volume}
  {102}},\ \bibinfo {pages} {057403} (\bibinfo {year} {2009})}\BibitemShut
  {NoStop}%
\bibitem [{\citenamefont {Steiner}\ \emph {et~al.}(2010)\citenamefont
  {Steiner}, \citenamefont {Neumann}, \citenamefont {Beck}, \citenamefont
  {Jelezko},\ and\ \citenamefont {Wrachtrup}}]{Steiner2010a}%
  \BibitemOpen
  \bibfield  {author} {\bibinfo {author} {\bibfnamefont {M.}~\bibnamefont
  {Steiner}}, \bibinfo {author} {\bibfnamefont {P.}~\bibnamefont {Neumann}},
  \bibinfo {author} {\bibfnamefont {J.}~\bibnamefont {Beck}}, \bibinfo {author}
  {\bibfnamefont {F.}~\bibnamefont {Jelezko}},\ and\ \bibinfo {author}
  {\bibfnamefont {J.}~\bibnamefont {Wrachtrup}},\ }\href
  {https://doi.org/10.1103/PhysRevB.81.035205} {\bibfield  {journal} {\bibinfo
  {journal} {Physical Review B}\ }\textbf {\bibinfo {volume} {81}},\ \bibinfo
  {pages} {035205} (\bibinfo {year} {2010})}\BibitemShut {NoStop}%
\bibitem [{\citenamefont {Wickenbrock}\ \emph {et~al.}(2016)\citenamefont
  {Wickenbrock}, \citenamefont {Zheng}, \citenamefont {Bougas}, \citenamefont
  {Leefer}, \citenamefont {Afach}, \citenamefont {Jarmola}, \citenamefont
  {Acosta},\ and\ \citenamefont {Budker}}]{Wickenbrock2016a}%
  \BibitemOpen
  \bibfield  {author} {\bibinfo {author} {\bibfnamefont {A.}~\bibnamefont
  {Wickenbrock}}, \bibinfo {author} {\bibfnamefont {H.}~\bibnamefont {Zheng}},
  \bibinfo {author} {\bibfnamefont {L.}~\bibnamefont {Bougas}}, \bibinfo
  {author} {\bibfnamefont {N.}~\bibnamefont {Leefer}}, \bibinfo {author}
  {\bibfnamefont {S.}~\bibnamefont {Afach}}, \bibinfo {author} {\bibfnamefont
  {A.}~\bibnamefont {Jarmola}}, \bibinfo {author} {\bibfnamefont {V.~M.}\
  \bibnamefont {Acosta}},\ and\ \bibinfo {author} {\bibfnamefont
  {D.}~\bibnamefont {Budker}},\ }\href {https://doi.org/10.1063/1.4960171}
  {\bibfield  {journal} {\bibinfo  {journal} {Applied Physics Letters}\
  }\textbf {\bibinfo {volume} {109}},\ \bibinfo {pages} {053505} (\bibinfo
  {year} {2016})}\BibitemShut {NoStop}%
\bibitem [{\citenamefont {Zheng}\ \emph {et~al.}(2020)\citenamefont {Zheng},
  \citenamefont {Sun}, \citenamefont {Chatzidrosos}, \citenamefont {Zhang},
  \citenamefont {Nakamura}, \citenamefont {Sumiya}, \citenamefont {Ohshima},
  \citenamefont {Isoya}, \citenamefont {Wrachtrup}, \citenamefont
  {Wickenbrock},\ and\ \citenamefont {Budker}}]{Zheng2020a}%
  \BibitemOpen
  \bibfield  {author} {\bibinfo {author} {\bibfnamefont {H.}~\bibnamefont
  {Zheng}}, \bibinfo {author} {\bibfnamefont {Z.}~\bibnamefont {Sun}}, \bibinfo
  {author} {\bibfnamefont {G.}~\bibnamefont {Chatzidrosos}}, \bibinfo {author}
  {\bibfnamefont {C.}~\bibnamefont {Zhang}}, \bibinfo {author} {\bibfnamefont
  {K.}~\bibnamefont {Nakamura}}, \bibinfo {author} {\bibfnamefont
  {H.}~\bibnamefont {Sumiya}}, \bibinfo {author} {\bibfnamefont
  {T.}~\bibnamefont {Ohshima}}, \bibinfo {author} {\bibfnamefont
  {J.}~\bibnamefont {Isoya}}, \bibinfo {author} {\bibfnamefont
  {J.}~\bibnamefont {Wrachtrup}}, \bibinfo {author} {\bibfnamefont
  {A.}~\bibnamefont {Wickenbrock}},\ and\ \bibinfo {author} {\bibfnamefont
  {D.}~\bibnamefont {Budker}},\ }\href
  {https://doi.org/10.1103/PhysRevApplied.13.044023} {\bibfield  {journal}
  {\bibinfo  {journal} {Physical Review Applied}\ }\textbf {\bibinfo {volume}
  {13}},\ \bibinfo {pages} {044023} (\bibinfo {year} {2020})}\BibitemShut
  {NoStop}%
\bibitem [{\citenamefont {Gali}(2009)}]{Gali2009a}%
  \BibitemOpen
  \bibfield  {author} {\bibinfo {author} {\bibfnamefont {A.}~\bibnamefont
  {Gali}},\ }\href {https://doi.org/10.1103/PhysRevB.80.241204} {\bibfield
  {journal} {\bibinfo  {journal} {Physical Review B}\ }\textbf {\bibinfo
  {volume} {80}},\ \bibinfo {pages} {241204} (\bibinfo {year}
  {2009})}\BibitemShut {NoStop}%
\bibitem [{\citenamefont {Felton}\ \emph {et~al.}(2009)\citenamefont {Felton},
  \citenamefont {Edmonds}, \citenamefont {Newton}, \citenamefont {Martineau},
  \citenamefont {Fisher}, \citenamefont {Twitchen},\ and\ \citenamefont
  {Baker}}]{Felton2009a}%
  \BibitemOpen
  \bibfield  {author} {\bibinfo {author} {\bibfnamefont {S.}~\bibnamefont
  {Felton}}, \bibinfo {author} {\bibfnamefont {A.~M.}\ \bibnamefont {Edmonds}},
  \bibinfo {author} {\bibfnamefont {M.~E.}\ \bibnamefont {Newton}}, \bibinfo
  {author} {\bibfnamefont {P.~M.}\ \bibnamefont {Martineau}}, \bibinfo {author}
  {\bibfnamefont {D.}~\bibnamefont {Fisher}}, \bibinfo {author} {\bibfnamefont
  {D.~J.}\ \bibnamefont {Twitchen}},\ and\ \bibinfo {author} {\bibfnamefont
  {J.~M.}\ \bibnamefont {Baker}},\ }\href
  {https://doi.org/10.1103/PhysRevB.79.075203} {\bibfield  {journal} {\bibinfo
  {journal} {Physical Review B}\ }\textbf {\bibinfo {volume} {79}},\ \bibinfo
  {pages} {075203} (\bibinfo {year} {2009})}\BibitemShut {NoStop}%
\bibitem [{\citenamefont {Cohen-Tannoudji}\ \emph {et~al.}(1998)\citenamefont
  {Cohen-Tannoudji}, \citenamefont {Dupont-Roc},\ and\ \citenamefont
  {Grynberg}}]{CohenTannoudji1998a}%
  \BibitemOpen
  \bibfield  {author} {\bibinfo {author} {\bibfnamefont {C.}~\bibnamefont
  {Cohen-Tannoudji}}, \bibinfo {author} {\bibfnamefont {J.}~\bibnamefont
  {Dupont-Roc}},\ and\ \bibinfo {author} {\bibfnamefont {G.}~\bibnamefont
  {Grynberg}},\ }\href@noop {} {\emph {\bibinfo {title} {Atom-{Photon}
  {Interactions}: {Basic} {Processes} and {Applications}}}}\ (\bibinfo
  {publisher} {Wiley},\ \bibinfo {year} {1998})\ \bibinfo {note} {, complement
  B1, page 45}\BibitemShut {NoStop}%
\bibitem [{\citenamefont {Childress}\ \emph {et~al.}(2006)\citenamefont
  {Childress}, \citenamefont {Gurudev~Dutt}, \citenamefont {Taylor},
  \citenamefont {Zibrov}, \citenamefont {Jelezko}, \citenamefont {Wrachtrup},
  \citenamefont {Hemmer},\ and\ \citenamefont {Lukin}}]{Childress2006a}%
  \BibitemOpen
  \bibfield  {author} {\bibinfo {author} {\bibfnamefont {L.}~\bibnamefont
  {Childress}}, \bibinfo {author} {\bibfnamefont {M.~V.}\ \bibnamefont
  {Gurudev~Dutt}}, \bibinfo {author} {\bibfnamefont {J.~M.}\ \bibnamefont
  {Taylor}}, \bibinfo {author} {\bibfnamefont {A.~S.}\ \bibnamefont {Zibrov}},
  \bibinfo {author} {\bibfnamefont {F.}~\bibnamefont {Jelezko}}, \bibinfo
  {author} {\bibfnamefont {J.}~\bibnamefont {Wrachtrup}}, \bibinfo {author}
  {\bibfnamefont {P.~R.}\ \bibnamefont {Hemmer}},\ and\ \bibinfo {author}
  {\bibfnamefont {M.~D.}\ \bibnamefont {Lukin}},\ }\href
  {https://doi.org/10.1126/science.1131871} {\bibfield  {journal} {\bibinfo
  {journal} {Science}\ }\textbf {\bibinfo {volume} {314}},\ \bibinfo {pages}
  {281} (\bibinfo {year} {2006})}\BibitemShut {NoStop}%
\bibitem [{\citenamefont {Chen}\ \emph {et~al.}(2015)\citenamefont {Chen},
  \citenamefont {Hirose},\ and\ \citenamefont {Cappellaro}}]{Chen2015a}%
  \BibitemOpen
  \bibfield  {author} {\bibinfo {author} {\bibfnamefont {M.}~\bibnamefont
  {Chen}}, \bibinfo {author} {\bibfnamefont {M.}~\bibnamefont {Hirose}},\ and\
  \bibinfo {author} {\bibfnamefont {P.}~\bibnamefont {Cappellaro}},\ }\href
  {https://doi.org/10.1103/PhysRevB.92.020101} {\bibfield  {journal} {\bibinfo
  {journal} {Physical Review B}\ }\textbf {\bibinfo {volume} {92}},\ \bibinfo
  {pages} {020101} (\bibinfo {year} {2015})}\BibitemShut {NoStop}%
\bibitem [{\citenamefont {Sangtawesin}\ \emph {et~al.}(2016)\citenamefont
  {Sangtawesin}, \citenamefont {McLellan}, \citenamefont {Myers}, \citenamefont
  {Jayich}, \citenamefont {Awschalom},\ and\ \citenamefont
  {Petta}}]{Sangtawesin2016a}%
  \BibitemOpen
  \bibfield  {author} {\bibinfo {author} {\bibfnamefont {S.}~\bibnamefont
  {Sangtawesin}}, \bibinfo {author} {\bibfnamefont {C.~A.}\ \bibnamefont
  {McLellan}}, \bibinfo {author} {\bibfnamefont {B.~A.}\ \bibnamefont {Myers}},
  \bibinfo {author} {\bibfnamefont {A.~C.~B.}\ \bibnamefont {Jayich}}, \bibinfo
  {author} {\bibfnamefont {D.~D.}\ \bibnamefont {Awschalom}},\ and\ \bibinfo
  {author} {\bibfnamefont {J.~R.}\ \bibnamefont {Petta}},\ }\href
  {https://doi.org/10.1088/1367-2630/18/8/083016} {\bibfield  {journal}
  {\bibinfo  {journal} {New Journal of Physics}\ }\textbf {\bibinfo {volume}
  {18}},\ \bibinfo {pages} {083016} (\bibinfo {year} {2016})}\BibitemShut
  {NoStop}%
\bibitem [{\citenamefont {Metsch}\ \emph {et~al.}(2019)\citenamefont {Metsch},
  \citenamefont {Senkalla}, \citenamefont {Tratzmiller}, \citenamefont
  {Scheuer}, \citenamefont {Kern}, \citenamefont {Achard}, \citenamefont
  {Tallaire}, \citenamefont {Plenio}, \citenamefont {Siyushev},\ and\
  \citenamefont {Jelezko}}]{Metsch2019a}%
  \BibitemOpen
  \bibfield  {author} {\bibinfo {author} {\bibfnamefont {M.~H.}\ \bibnamefont
  {Metsch}}, \bibinfo {author} {\bibfnamefont {K.}~\bibnamefont {Senkalla}},
  \bibinfo {author} {\bibfnamefont {B.}~\bibnamefont {Tratzmiller}}, \bibinfo
  {author} {\bibfnamefont {J.}~\bibnamefont {Scheuer}}, \bibinfo {author}
  {\bibfnamefont {M.}~\bibnamefont {Kern}}, \bibinfo {author} {\bibfnamefont
  {J.}~\bibnamefont {Achard}}, \bibinfo {author} {\bibfnamefont
  {A.}~\bibnamefont {Tallaire}}, \bibinfo {author} {\bibfnamefont {M.~B.}\
  \bibnamefont {Plenio}}, \bibinfo {author} {\bibfnamefont {P.}~\bibnamefont
  {Siyushev}},\ and\ \bibinfo {author} {\bibfnamefont {F.}~\bibnamefont
  {Jelezko}},\ }\href {https://doi.org/10.1103/PhysRevLett.122.190503}
  {\bibfield  {journal} {\bibinfo  {journal} {Physical Review Letters}\
  }\textbf {\bibinfo {volume} {122}},\ \bibinfo {pages} {190503} (\bibinfo
  {year} {2019})}\BibitemShut {NoStop}%
\bibitem [{\citenamefont {Broadway}\ \emph {et~al.}(2018)\citenamefont
  {Broadway}, \citenamefont {Lillie}, \citenamefont {Dontschuk}, \citenamefont
  {Stacey}, \citenamefont {Hall}, \citenamefont {Tetienne},\ and\ \citenamefont
  {Hollenberg}}]{Broadway2018a}%
  \BibitemOpen
  \bibfield  {author} {\bibinfo {author} {\bibfnamefont {D.~A.}\ \bibnamefont
  {Broadway}}, \bibinfo {author} {\bibfnamefont {S.~E.}\ \bibnamefont
  {Lillie}}, \bibinfo {author} {\bibfnamefont {N.}~\bibnamefont {Dontschuk}},
  \bibinfo {author} {\bibfnamefont {A.}~\bibnamefont {Stacey}}, \bibinfo
  {author} {\bibfnamefont {L.~T.}\ \bibnamefont {Hall}}, \bibinfo {author}
  {\bibfnamefont {J.-P.}\ \bibnamefont {Tetienne}},\ and\ \bibinfo {author}
  {\bibfnamefont {L.~C.~L.}\ \bibnamefont {Hollenberg}},\ }\href
  {https://doi.org/10.1063/1.5021491} {\bibfield  {journal} {\bibinfo
  {journal} {Applied Physics Letters}\ }\textbf {\bibinfo {volume} {112}},\
  \bibinfo {pages} {103103} (\bibinfo {year} {2018})}\BibitemShut {NoStop}%
\bibitem [{\citenamefont {Lesik}\ \emph {et~al.}(2015)\citenamefont {Lesik},
  \citenamefont {Plays}, \citenamefont {Tallaire}, \citenamefont {Achard},
  \citenamefont {Brinza}, \citenamefont {William}, \citenamefont {Chipaux},
  \citenamefont {Toraille}, \citenamefont {Debuisschert}, \citenamefont
  {Gicquel}, \citenamefont {Roch},\ and\ \citenamefont {Jacques}}]{Lesik2015a}%
  \BibitemOpen
  \bibfield  {author} {\bibinfo {author} {\bibfnamefont {M.}~\bibnamefont
  {Lesik}}, \bibinfo {author} {\bibfnamefont {T.}~\bibnamefont {Plays}},
  \bibinfo {author} {\bibfnamefont {A.}~\bibnamefont {Tallaire}}, \bibinfo
  {author} {\bibfnamefont {J.}~\bibnamefont {Achard}}, \bibinfo {author}
  {\bibfnamefont {O.}~\bibnamefont {Brinza}}, \bibinfo {author} {\bibfnamefont
  {L.}~\bibnamefont {William}}, \bibinfo {author} {\bibfnamefont
  {M.}~\bibnamefont {Chipaux}}, \bibinfo {author} {\bibfnamefont
  {L.}~\bibnamefont {Toraille}}, \bibinfo {author} {\bibfnamefont
  {T.}~\bibnamefont {Debuisschert}}, \bibinfo {author} {\bibfnamefont
  {A.}~\bibnamefont {Gicquel}}, \bibinfo {author} {\bibfnamefont {J.~F.}\
  \bibnamefont {Roch}},\ and\ \bibinfo {author} {\bibfnamefont
  {V.}~\bibnamefont {Jacques}},\ }\href
  {https://doi.org/10.1016/j.diamond.2015.05.003} {\bibfield  {journal}
  {\bibinfo  {journal} {Diamond and Related Materials}\ }\textbf {\bibinfo
  {volume} {56}},\ \bibinfo {pages} {47} (\bibinfo {year} {2015})}\BibitemShut
  {NoStop}%
\bibitem [{\citenamefont {Budker}\ and\ \citenamefont
  {Romalis}(2007)}]{Budker2007a}%
  \BibitemOpen
  \bibfield  {author} {\bibinfo {author} {\bibfnamefont {D.}~\bibnamefont
  {Budker}}\ and\ \bibinfo {author} {\bibfnamefont {M.}~\bibnamefont
  {Romalis}},\ }\href {https://doi.org/10.1038/nphys566} {\bibfield  {journal}
  {\bibinfo  {journal} {Nature Physics}\ }\textbf {\bibinfo {volume} {3}},\
  \bibinfo {pages} {227} (\bibinfo {year} {2007})}\BibitemShut {NoStop}%
\bibitem [{\citenamefont {Barry}\ \emph {et~al.}(2020)\citenamefont {Barry},
  \citenamefont {Schloss}, \citenamefont {Bauch}, \citenamefont {Turner},
  \citenamefont {Hart}, \citenamefont {Pham},\ and\ \citenamefont
  {Walsworth}}]{Barry2020a}%
  \BibitemOpen
  \bibfield  {author} {\bibinfo {author} {\bibfnamefont {J.~F.}\ \bibnamefont
  {Barry}}, \bibinfo {author} {\bibfnamefont {J.~M.}\ \bibnamefont {Schloss}},
  \bibinfo {author} {\bibfnamefont {E.}~\bibnamefont {Bauch}}, \bibinfo
  {author} {\bibfnamefont {M.~J.}\ \bibnamefont {Turner}}, \bibinfo {author}
  {\bibfnamefont {C.~A.}\ \bibnamefont {Hart}}, \bibinfo {author}
  {\bibfnamefont {L.~M.}\ \bibnamefont {Pham}},\ and\ \bibinfo {author}
  {\bibfnamefont {R.~L.}\ \bibnamefont {Walsworth}},\ }\href
  {https://doi.org/10.1103/RevModPhys.92.015004} {\bibfield  {journal}
  {\bibinfo  {journal} {Reviews of Modern Physics}\ }\textbf {\bibinfo {volume}
  {92}},\ \bibinfo {pages} {015004} (\bibinfo {year} {2020})}\BibitemShut
  {NoStop}%
\bibitem [{\citenamefont {Chatzidrosos}(2021)}]{Chatzidrosos2021a}%
  \BibitemOpen
  \bibfield  {author} {\bibinfo {author} {\bibfnamefont {G.}~\bibnamefont
  {Chatzidrosos}},\ }\emph {\bibinfo {title} {A perfect imperfection: {Quantum}
  magnetometry and applications using nitrogen-vacancy defects in diamond}},\
  \href {https://openscience.ub.uni-mainz.de/handle/20.500.12030/6372} {Ph.D.
  thesis},\ \bibinfo  {school} {Johannes Gutenberg-Universit{\"a}t Mainz}
  (\bibinfo {year} {2021})\BibitemShut {NoStop}%
\bibitem [{\citenamefont {Zhang}\ \emph {et~al.}(2021)\citenamefont {Zhang},
  \citenamefont {Shagieva}, \citenamefont {Widmann}, \citenamefont
  {K{\"u}bler}, \citenamefont {Vorobyov}, \citenamefont {Kapitanova},
  \citenamefont {Nenasheva}, \citenamefont {Corkill}, \citenamefont {Rhrle},
  \citenamefont {Nakamura}, \citenamefont {Sumiya}, \citenamefont {Onoda},
  \citenamefont {Isoya},\ and\ \citenamefont {Wrachtrup}}]{Zhang2021a}%
  \BibitemOpen
  \bibfield  {author} {\bibinfo {author} {\bibfnamefont {C.}~\bibnamefont
  {Zhang}}, \bibinfo {author} {\bibfnamefont {F.}~\bibnamefont {Shagieva}},
  \bibinfo {author} {\bibfnamefont {M.}~\bibnamefont {Widmann}}, \bibinfo
  {author} {\bibfnamefont {M.}~\bibnamefont {K{\"u}bler}}, \bibinfo {author}
  {\bibfnamefont {V.}~\bibnamefont {Vorobyov}}, \bibinfo {author}
  {\bibfnamefont {P.}~\bibnamefont {Kapitanova}}, \bibinfo {author}
  {\bibfnamefont {E.}~\bibnamefont {Nenasheva}}, \bibinfo {author}
  {\bibfnamefont {R.}~\bibnamefont {Corkill}}, \bibinfo {author} {\bibfnamefont
  {O.}~\bibnamefont {Rhrle}}, \bibinfo {author} {\bibfnamefont
  {K.}~\bibnamefont {Nakamura}}, \bibinfo {author} {\bibfnamefont
  {H.}~\bibnamefont {Sumiya}}, \bibinfo {author} {\bibfnamefont
  {S.}~\bibnamefont {Onoda}}, \bibinfo {author} {\bibfnamefont
  {J.}~\bibnamefont {Isoya}},\ and\ \bibinfo {author} {\bibfnamefont
  {J.}~\bibnamefont {Wrachtrup}},\ }\href
  {https://doi.org/10.1103/PhysRevApplied.15.064075} {\bibfield  {journal}
  {\bibinfo  {journal} {Physical Review Applied}\ }\textbf {\bibinfo {volume}
  {15}},\ \bibinfo {pages} {064075} (\bibinfo {year} {2021})}\BibitemShut
  {NoStop}%
\bibitem [{\citenamefont {Patel}\ \emph {et~al.}(2020)\citenamefont {Patel},
  \citenamefont {Zhou}, \citenamefont {Frangeskou}, \citenamefont {Stimpson},
  \citenamefont {Breeze}, \citenamefont {Nikitin}, \citenamefont {Dale},
  \citenamefont {Nichols}, \citenamefont {Thornley}, \citenamefont {Green},
  \citenamefont {Newton}, \citenamefont {Edmonds}, \citenamefont {Markham},
  \citenamefont {Twitchen},\ and\ \citenamefont {Morley}}]{Patel2020a}%
  \BibitemOpen
  \bibfield  {author} {\bibinfo {author} {\bibfnamefont {R.}~\bibnamefont
  {Patel}}, \bibinfo {author} {\bibfnamefont {L.}~\bibnamefont {Zhou}},
  \bibinfo {author} {\bibfnamefont {A.}~\bibnamefont {Frangeskou}}, \bibinfo
  {author} {\bibfnamefont {G.}~\bibnamefont {Stimpson}}, \bibinfo {author}
  {\bibfnamefont {B.}~\bibnamefont {Breeze}}, \bibinfo {author} {\bibfnamefont
  {A.}~\bibnamefont {Nikitin}}, \bibinfo {author} {\bibfnamefont
  {M.}~\bibnamefont {Dale}}, \bibinfo {author} {\bibfnamefont {E.}~\bibnamefont
  {Nichols}}, \bibinfo {author} {\bibfnamefont {W.}~\bibnamefont {Thornley}},
  \bibinfo {author} {\bibfnamefont {B.}~\bibnamefont {Green}}, \bibinfo
  {author} {\bibfnamefont {M.}~\bibnamefont {Newton}}, \bibinfo {author}
  {\bibfnamefont {A.}~\bibnamefont {Edmonds}}, \bibinfo {author} {\bibfnamefont
  {M.}~\bibnamefont {Markham}}, \bibinfo {author} {\bibfnamefont
  {D.}~\bibnamefont {Twitchen}},\ and\ \bibinfo {author} {\bibfnamefont
  {G.}~\bibnamefont {Morley}},\ }\href
  {https://doi.org/10.1103/PhysRevApplied.14.044058} {\bibfield  {journal}
  {\bibinfo  {journal} {Physical Review Applied}\ }\textbf {\bibinfo {volume}
  {14}},\ \bibinfo {pages} {044058} (\bibinfo {year} {2020})}\BibitemShut
  {NoStop}%
\bibitem [{\citenamefont {Shields}\ \emph {et~al.}(2015)\citenamefont
  {Shields}, \citenamefont {Unterreithmeier}, \citenamefont {de~Leon},
  \citenamefont {Park},\ and\ \citenamefont {Lukin}}]{Shields2015a}%
  \BibitemOpen
  \bibfield  {author} {\bibinfo {author} {\bibfnamefont {B.}~\bibnamefont
  {Shields}}, \bibinfo {author} {\bibfnamefont {Q.}~\bibnamefont
  {Unterreithmeier}}, \bibinfo {author} {\bibfnamefont {N.}~\bibnamefont
  {de~Leon}}, \bibinfo {author} {\bibfnamefont {H.}~\bibnamefont {Park}},\ and\
  \bibinfo {author} {\bibfnamefont {M.}~\bibnamefont {Lukin}},\ }\href
  {https://doi.org/10.1103/PhysRevLett.114.136402} {\bibfield  {journal}
  {\bibinfo  {journal} {Physical Review Letters}\ }\textbf {\bibinfo {volume}
  {114}},\ \bibinfo {pages} {136402} (\bibinfo {year} {2015})}\BibitemShut
  {NoStop}%
\bibitem [{\citenamefont {Hopper}\ \emph {et~al.}(2018)\citenamefont {Hopper},
  \citenamefont {Grote}, \citenamefont {Parks},\ and\ \citenamefont
  {Bassett}}]{Hopper2018a}%
  \BibitemOpen
  \bibfield  {author} {\bibinfo {author} {\bibfnamefont {D.~A.}\ \bibnamefont
  {Hopper}}, \bibinfo {author} {\bibfnamefont {R.~R.}\ \bibnamefont {Grote}},
  \bibinfo {author} {\bibfnamefont {S.~M.}\ \bibnamefont {Parks}},\ and\
  \bibinfo {author} {\bibfnamefont {L.~C.}\ \bibnamefont {Bassett}},\ }\href
  {https://doi.org/10.1021/acsnano.8b01265} {\bibfield  {journal} {\bibinfo
  {journal} {ACS Nano}\ }\textbf {\bibinfo {volume} {12}},\ \bibinfo {pages}
  {4678} (\bibinfo {year} {2018})}\BibitemShut {NoStop}%
\bibitem [{\citenamefont {Chu}\ \emph {et~al.}(2014)\citenamefont {Chu},
  \citenamefont {de~Leon}, \citenamefont {Shields}, \citenamefont {Hausmann},
  \citenamefont {Evans}, \citenamefont {Togan}, \citenamefont {Burek},
  \citenamefont {Markham}, \citenamefont {Stacey}, \citenamefont {Zibrov},
  \citenamefont {Yacoby}, \citenamefont {Twitchen}, \citenamefont {Loncar},
  \citenamefont {Park}, \citenamefont {Maletinsky},\ and\ \citenamefont
  {Lukin}}]{Chu2014a}%
  \BibitemOpen
  \bibfield  {author} {\bibinfo {author} {\bibfnamefont {Y.}~\bibnamefont
  {Chu}}, \bibinfo {author} {\bibfnamefont {N.}~\bibnamefont {de~Leon}},
  \bibinfo {author} {\bibfnamefont {B.}~\bibnamefont {Shields}}, \bibinfo
  {author} {\bibfnamefont {B.}~\bibnamefont {Hausmann}}, \bibinfo {author}
  {\bibfnamefont {R.}~\bibnamefont {Evans}}, \bibinfo {author} {\bibfnamefont
  {E.}~\bibnamefont {Togan}}, \bibinfo {author} {\bibfnamefont {M.~J.}\
  \bibnamefont {Burek}}, \bibinfo {author} {\bibfnamefont {M.}~\bibnamefont
  {Markham}}, \bibinfo {author} {\bibfnamefont {A.}~\bibnamefont {Stacey}},
  \bibinfo {author} {\bibfnamefont {A.}~\bibnamefont {Zibrov}}, \bibinfo
  {author} {\bibfnamefont {A.}~\bibnamefont {Yacoby}}, \bibinfo {author}
  {\bibfnamefont {D.}~\bibnamefont {Twitchen}}, \bibinfo {author}
  {\bibfnamefont {M.}~\bibnamefont {Loncar}}, \bibinfo {author} {\bibfnamefont
  {H.}~\bibnamefont {Park}}, \bibinfo {author} {\bibfnamefont {P.}~\bibnamefont
  {Maletinsky}},\ and\ \bibinfo {author} {\bibfnamefont {M.}~\bibnamefont
  {Lukin}},\ }\href {https://doi.org/10.1021/nl404836p} {\bibfield  {journal}
  {\bibinfo  {journal} {Nano Letters}\ }\textbf {\bibinfo {volume} {14}},\
  \bibinfo {pages} {1982} (\bibinfo {year} {2014})}\BibitemShut {NoStop}%
\bibitem [{\citenamefont {Pezzagna}\ \emph {et~al.}(2010)\citenamefont
  {Pezzagna}, \citenamefont {Naydenov}, \citenamefont {Jelezko}, \citenamefont
  {Wrachtrup},\ and\ \citenamefont {Meijer}}]{Pezzagna2010a}%
  \BibitemOpen
  \bibfield  {author} {\bibinfo {author} {\bibfnamefont {S.}~\bibnamefont
  {Pezzagna}}, \bibinfo {author} {\bibfnamefont {B.}~\bibnamefont {Naydenov}},
  \bibinfo {author} {\bibfnamefont {F.}~\bibnamefont {Jelezko}}, \bibinfo
  {author} {\bibfnamefont {J.}~\bibnamefont {Wrachtrup}},\ and\ \bibinfo
  {author} {\bibfnamefont {J.}~\bibnamefont {Meijer}},\ }\href
  {https://doi.org/10.1088/1367-2630/12/6/065017} {\bibfield  {journal}
  {\bibinfo  {journal} {New Journal of Physics}\ }\textbf {\bibinfo {volume}
  {12}},\ \bibinfo {pages} {065017} (\bibinfo {year} {2010})}\BibitemShut
  {NoStop}%
\bibitem [{\citenamefont {Hedrich}\ \emph {et~al.}(2020)\citenamefont
  {Hedrich}, \citenamefont {Rohner}, \citenamefont {Batzer}, \citenamefont
  {Maletinsky},\ and\ \citenamefont {Shields}}]{Hedrich2020a}%
  \BibitemOpen
  \bibfield  {author} {\bibinfo {author} {\bibfnamefont {N.}~\bibnamefont
  {Hedrich}}, \bibinfo {author} {\bibfnamefont {D.}~\bibnamefont {Rohner}},
  \bibinfo {author} {\bibfnamefont {M.}~\bibnamefont {Batzer}}, \bibinfo
  {author} {\bibfnamefont {P.}~\bibnamefont {Maletinsky}},\ and\ \bibinfo
  {author} {\bibfnamefont {B.~J.}\ \bibnamefont {Shields}},\ }\href
  {https://doi.org/10.1103/PhysRevApplied.14.064007} {\bibfield  {journal}
  {\bibinfo  {journal} {Physical Review Applied}\ }\textbf {\bibinfo {volume}
  {14}},\ \bibinfo {pages} {064007} (\bibinfo {year} {2020})}\BibitemShut
  {NoStop}%
\bibitem [{\citenamefont {Balasubramanian}\ \emph {et~al.}(2022)\citenamefont
  {Balasubramanian}, \citenamefont {Osterkamp}, \citenamefont {Brinza},
  \citenamefont {Rollo}, \citenamefont {Robert-Philip}, \citenamefont
  {Goldner}, \citenamefont {Jacques}, \citenamefont {Jelezko}, \citenamefont
  {Achard},\ and\ \citenamefont {Tallaire}}]{Balasubramanian2022a}%
  \BibitemOpen
  \bibfield  {author} {\bibinfo {author} {\bibfnamefont {P.}~\bibnamefont
  {Balasubramanian}}, \bibinfo {author} {\bibfnamefont {C.}~\bibnamefont
  {Osterkamp}}, \bibinfo {author} {\bibfnamefont {O.}~\bibnamefont {Brinza}},
  \bibinfo {author} {\bibfnamefont {M.}~\bibnamefont {Rollo}}, \bibinfo
  {author} {\bibfnamefont {I.}~\bibnamefont {Robert-Philip}}, \bibinfo {author}
  {\bibfnamefont {P.}~\bibnamefont {Goldner}}, \bibinfo {author} {\bibfnamefont
  {V.}~\bibnamefont {Jacques}}, \bibinfo {author} {\bibfnamefont
  {F.}~\bibnamefont {Jelezko}}, \bibinfo {author} {\bibfnamefont
  {J.}~\bibnamefont {Achard}},\ and\ \bibinfo {author} {\bibfnamefont
  {A.}~\bibnamefont {Tallaire}},\ }\href
  {https://doi.org/10.1016/j.carbon.2022.04.005} {\bibfield  {journal}
  {\bibinfo  {journal} {Carbon}\ }\textbf {\bibinfo {volume} {194}},\ \bibinfo
  {pages} {282} (\bibinfo {year} {2022})}\BibitemShut {NoStop}%
\bibitem [{\citenamefont {Knecht}\ \emph {et~al.}(2016)\citenamefont {Knecht},
  \citenamefont {Pravdivtsev}, \citenamefont {H{\"o}vener}, \citenamefont
  {Yurkovskaya},\ and\ \citenamefont {Ivanov}}]{Knecht2016a}%
  \BibitemOpen
  \bibfield  {author} {\bibinfo {author} {\bibfnamefont {S.}~\bibnamefont
  {Knecht}}, \bibinfo {author} {\bibfnamefont {A.~N.}\ \bibnamefont
  {Pravdivtsev}}, \bibinfo {author} {\bibfnamefont {J.-B.}\ \bibnamefont
  {H{\"o}vener}}, \bibinfo {author} {\bibfnamefont {A.~V.}\ \bibnamefont
  {Yurkovskaya}},\ and\ \bibinfo {author} {\bibfnamefont {K.~L.}\ \bibnamefont
  {Ivanov}},\ }\href {https://doi.org/10.1039/C5RA28059A} {\bibfield  {journal}
  {\bibinfo  {journal} {RSC Advances}\ }\textbf {\bibinfo {volume} {6}},\
  \bibinfo {pages} {24470} (\bibinfo {year} {2016})}\BibitemShut {NoStop}%
\bibitem [{\citenamefont {Manson}\ \emph {et~al.}(2006)\citenamefont {Manson},
  \citenamefont {Harrison},\ and\ \citenamefont {Sellars}}]{Manson2006a}%
  \BibitemOpen
  \bibfield  {author} {\bibinfo {author} {\bibfnamefont {N.~B.}\ \bibnamefont
  {Manson}}, \bibinfo {author} {\bibfnamefont {J.~P.}\ \bibnamefont
  {Harrison}},\ and\ \bibinfo {author} {\bibfnamefont {M.~J.}\ \bibnamefont
  {Sellars}},\ }\href {https://doi.org/10.1103/PhysRevB.74.104303} {\bibfield
  {journal} {\bibinfo  {journal} {Physical Review B}\ }\textbf {\bibinfo
  {volume} {74}},\ \bibinfo {pages} {104303} (\bibinfo {year}
  {2006})}\BibitemShut {NoStop}%
\bibitem [{\citenamefont {Robledo}\ \emph {et~al.}(2011)\citenamefont
  {Robledo}, \citenamefont {Bernien}, \citenamefont {Sar},\ and\ \citenamefont
  {Hanson}}]{Robledo2011a}%
  \BibitemOpen
  \bibfield  {author} {\bibinfo {author} {\bibfnamefont {L.}~\bibnamefont
  {Robledo}}, \bibinfo {author} {\bibfnamefont {H.}~\bibnamefont {Bernien}},
  \bibinfo {author} {\bibfnamefont {T.~v.~d.}\ \bibnamefont {Sar}},\ and\
  \bibinfo {author} {\bibfnamefont {R.}~\bibnamefont {Hanson}},\ }\href
  {https://doi.org/10.1088/1367-2630/13/2/025013} {\bibfield  {journal}
  {\bibinfo  {journal} {New Journal of Physics}\ }\textbf {\bibinfo {volume}
  {13}},\ \bibinfo {pages} {025013} (\bibinfo {year} {2011})}\BibitemShut
  {NoStop}%
\bibitem [{\citenamefont {Gupta}\ \emph {et~al.}(2016)\citenamefont {Gupta},
  \citenamefont {Hacquebard},\ and\ \citenamefont {Childress}}]{Gupta2016a}%
  \BibitemOpen
  \bibfield  {author} {\bibinfo {author} {\bibfnamefont {A.}~\bibnamefont
  {Gupta}}, \bibinfo {author} {\bibfnamefont {L.}~\bibnamefont {Hacquebard}},\
  and\ \bibinfo {author} {\bibfnamefont {L.}~\bibnamefont {Childress}},\ }\href
  {https://doi.org/10.1364/JOSAB.33.000B28} {\bibfield  {journal} {\bibinfo
  {journal} {JOSA B}\ }\textbf {\bibinfo {volume} {33}},\ \bibinfo {pages}
  {B28} (\bibinfo {year} {2016})}\BibitemShut {NoStop}%
\end{thebibliography}%

\newpage
\begin{widetext}
\newpage

	\section{Supplementary Information}

	\subsection{I. Effective Nuclear Hamiltonian}\label{SI1}

Here, we calculate an effective Hamiltonian for the $^{15}$N spin in the electronic $m_S=0$ subspace.  
To that end, we employ Van Vleck perturbation theory, which is applicable to Hamiltonians $\hat{H}$ that can be written in the form $\hat{H} = \hat{H}_0 + \hat{V}$, where $\hat{H}_0$ is block diagonal, made of distinct subspaces, and $V$ is a perturbation that couples the initially uncoupled different subspaces of $\hat{H}_0$. 
Following the notation in\,\cite{CohenTannoudji1998a}, to second order, the effective Hamiltonian for each individual subspace is given by
\begin{align}\nonumber
\bra{i}\hat{H}^{\alpha}_{\rm{eff}}\ket{j} =& \bra{i,\alpha}H_0+V\ket{j,\alpha} \\ \nonumber
&+\frac{1}{2}\sum_{k,\gamma\ne\alpha} \bra{i,\alpha}V\ket{k,\gamma}\bra{k,\gamma}V\ket{j,\alpha} \\
&\quad\times \left[ \frac{1}{E_{i,\alpha}-E_{k,\gamma}} + \frac{1}{E_{k,\alpha}-E_{k,\gamma}} \right].
\label{eq:vanvleck}
\end{align}
Here, latin indices denote states within a given subspace, and greek indicies count over the subspaces.  
Equation\,\eqref{eq:vanvleck} is valid if the energy difference between states in different blocks is much larger than the coupling between them, $|E_{i,\alpha} - E_{j,\beta}| \gg \bra{i,\alpha} V \ket{j,\beta} $.

Hamiltonian $\hat{H}^{\rm gs}$ from the main text is such a block diagonal Hamiltonian. Written in the basis $\{\ket{m_S,m_I}\}$, it reads
\begin{equation}
\tiny
\frac{\hat{H}^{\rm gs}}{h} = 
\left[
\begin{matrix}
D_0^{\rm gs} + \gamma_SB_z + \frac{\gamma_IB_z + A_\parallel^{\rm gs}}{2} & \frac{\gamma_I}{2}B_x & \frac{\gamma_S}{\sqrt{2}}B_x & 0 & 0 & 0 \\
\frac{\gamma_I}{2}B_x & D_0^{\rm gs} + \gamma_SB_z - \frac{\gamma_IB_z + A_\parallel^{\rm gs}}{2} & \frac{1}{\sqrt{2}}A_\perp^{\rm gs} & \frac{\gamma_S}{\sqrt{2}}B_x & 0 & 0 \\
\frac{\gamma_S}{\sqrt{2}}B_x & \frac{1}{\sqrt{2}}A_\perp^{\rm gs} & + \tfrac{1}{2}\gamma_IB_z & \frac{\gamma_I}{2}B_x & \frac{\gamma_S}{\sqrt{2}}B_x & 0 \\
0 & \frac{\gamma_S}{\sqrt{2}}B_x & \frac{\gamma_I}{2}B_x & - \tfrac{1}{2}\gamma_IB_z & \frac{1}{\sqrt{2}}A_\perp^{\rm gs} & \frac{\gamma_S}{\sqrt{2}}B_x \\
0 & 0 & \frac{\gamma_S}{\sqrt{2}}B_x & \frac{1}{\sqrt{2}}A_\perp^{\rm gs} & D_0^{\rm gs} - \gamma_SB_z + \frac{\gamma_IB_z - A_\parallel^{\rm gs}}{2} & \frac{\gamma_I}{2}B_x \\
0 & 0 & 0 & \frac{\gamma_S}{\sqrt{2}}B_x & \frac{\gamma_I}{2}B_x & D_0^{\rm gs} - \gamma_SB_z - \frac{\gamma_IB_z - A_\parallel^{\rm gs}}{2} \\
\end{matrix}
\right]
\normalsize
\end{equation}
where the blocks are defined by states of equal values of $m_S$, and where, without loss of generality, we define the direction of the transverse magnetic field as the $x$-direction, e.g. that $B_y=0$ and $B_x =: B_\perp$.

We now evaluate Eq.\,\eqref{eq:vanvleck} for $\hat{H}^{\rm gs} = \hat{H}_0 + \hat{V}$ with $\hat{H}_0=D_0^{\rm gs}\hat{S}_z^2 +  \gamma_{S}b_z\hat{S}_z + \gamma_{I}b_z\hat{I}_z$ and $\hat{V} = B_\perp(\gamma_I \hat{I}_x + \gamma_{S} \hat{S}_x) + \hat{\bm{S}}\cdot\bm{A}\cdot\hat{\bm{I}}$, and $\alpha$ corresponding to the $m_S=0$ subspace. This leads to the following matrix elements,
\begin{align}
\small
\frac{{\bra{1}\hat{H}_{\rm eff}^{m_S=0}}\ket{1}}{h} = \ &\frac{+\gamma_IB_z}{2} + \frac{(A_\perp^{\rm gs})^2}{A_\parallel^{\rm gs} - 2D_0^{\rm gs} + 2\gamma_IB_z - 2\gamma_SB_z} \nonumber\\
&+ \frac{4\gamma_S^2B_\perp^2D_0^{\rm gs}}{(A_\parallel^{\rm gs}-2D_0^{\rm gs}+2\gamma_SB_z)(A_\parallel^{\rm gs}+2D_0^{\rm gs}+2\gamma_SB_z)}\ ,\\
\frac{{\bra{2}\hat{H}_{\rm eff}^{m_S=0}}\ket{2}}{h} = \ &\frac{-\gamma_IB_z}{2} + \frac{(A_\perp^{\rm gs})^2}{A_\parallel^{\rm gs} - 2D_0^{\rm gs} - 2\gamma_IB_z + 2\gamma_SB_z} \nonumber\\
&+ \frac{4\gamma_S^2B_\perp^2D_0^{\rm gs}}{(A_\parallel^{\rm gs}-2D_0^{\rm gs}-2\gamma_SB_z)(A_\parallel^{\rm gs}+2D_0^{\rm gs}-2\gamma_SB_z)}\ ,\\
\frac{{\bra{1}\hat{H}_{\rm eff}^{m_S=0}}\ket{2}}{h} = \ &\frac{{\bra{2}\hat{H}_{\rm eff}^{m_S=0}}\ket{1}}{h} = \frac{+\gamma_IB_\perp}{2}
 + \frac{\gamma_SB_\perp A_\perp^{\rm gs}/2}{A_\parallel^{\rm gs}-2D_0^{\rm gs}+2\gamma_SB_z}
 + \frac{\gamma_SB_\perp A_\perp^{\rm gs}/2}{A_\parallel^{\rm gs}-2D_0^{\rm gs}-2\gamma_SB_z} \nonumber\\
&+ \frac{\gamma_SB_\perp A_\perp^{\rm gs}/2}{A_\parallel^{\rm gs}-2D_0^{\rm gs}+2\gamma_SB_z-2\gamma_IB_z}
 + \frac{\gamma_SB_\perp A_\perp^{\rm gs}/2}{A_\parallel^{\rm gs}-2D_0^{\rm gs}-2\gamma_SB_z+2\gamma_IB_z}\ .
\end{align}
Since $\gamma_I \ll \gamma_S$, we simplify in all denominators the terms $(\gamma_S\pm\gamma_I)\approx\gamma_S$. 
In similar fashion, we use the fact that $A_\parallel^{\rm gs} \ll D_0^{\rm gs}$ and thus set $(A_\parallel^{\rm gs} \pm 2D_0^{\rm gs})\approx \pm 2D_0^{\rm gs}$ in all denominators. 
We obtain the following effective Hamiltonian for the $^{15}$N spin in the $m_S=0$ manifold:
\begin{align}
\small
\frac{\hat{H}^{m_S=0}_{\rm{eff}}}{h}
& = \frac{1}{2}\begin{bmatrix} 
+ \gamma_I B_z +\frac{+\gamma_S B_z (A_\perp^{\rm gs})^2 - D_0^{\rm gs}(A_\perp^{\rm gs})^2 - 2D_0^{\rm gs}(\gamma_SB_\perp)^2}{(D_0^{\rm gs})^2 - (\gamma_SB_z)^2}
& \gamma_I B_\perp - \frac{2\gamma_S B_\perp A_\perp^{\rm gs}D_0^{\rm gs}}{(D_0^{\rm gs})^2 - (\gamma_SB_z)^2}\\
  \gamma_I B_\perp - \frac{2\gamma_S B_\perp A_\perp^{\rm gs}D_0^{\rm gs}}{(D_0^{\rm gs})^2 - (\gamma_SB_z)^2}
&-\gamma_I B_z +\frac{-\gamma_S B_z (A_\perp^{\rm gs})^2 - D_0^{\rm gs}(A_\perp^{\rm gs})^2 - 2D_0^{\rm gs}(\gamma_SB_\perp)^2}{(D_0^{\rm gs})^2 - (\gamma_SB_z)^2} \end{bmatrix}.
\end{align}
Next, we add an energy-offset of $\left(D_0^{\rm gs}(A_\perp^{\rm gs})^2 - 2D_0^{\rm gs}(\gamma_SB_\perp)^2\right)/\left((D_0^{\rm gs})^2 - (\gamma_SB_z)^2\right)$ to place the energy levels symmetrically around zero, and thereby obtain the Hamiltonian given in the main text,
\begin{gather}
\frac{\hat{H}^{m_S=0}_{\rm{eff}}}{h} = \frac{1}{2}\,\begin{bmatrix}\gamma_I B_z + \nu_z &\hspace{0.2cm} \gamma_I B_\perp + \nu_\perp \vspace{0.2cm}\\ \gamma_I B_\perp + \nu_\perp &\hspace{0.2cm} - \gamma_I B_z - \nu_z \end{bmatrix}.
\end{gather}
where
\begin{equation}
\nu_z = \frac{\gamma_SB_z(A_\perp^{\rm gs})^2}{(D_0^{\rm gs})^2-(\gamma_SB_z)^2}
\end{equation}
denotes the correction to the diagonal elements caused by mixing between states of different $m_S$, and 
\begin{equation}
\nu_\perp = \frac{-2\,\gamma_SB_\perp A_\perp^{\rm gs}D_0^{\rm gs}}{(D_0^{\rm gs})^2-(\gamma_SB_z)^2}
\end{equation}
is the corresponding correction to the off-diagonal elements.
Note that these expressions for $\nu_z$ and $\nu_\perp$ are diverging for $D_0^{\rm gs}=\gamma_SB_z$, e.g. near the ground state level anti-crossing.
However, Van Vleck formalism is not applicable to that regime since the corresponding electronic subspaces of $H_0$ are not sufficiently spaced in energy once this condition is approached.

	\subsection{II. Numerical Model for Optical Pumping under ESLAC Conditions}\label{SI2}

In this section, we present in detail our numerical model for simulating the spin dynamics of the NV center with and without green illumination for a given magnetic field $\bm{B}$.
To calculate the trajectory of both the electron and nuclear spin under optical pumping it is necessary to consider not only classical rate equations coupling the orbital states, but also to incorporate the quantum mechanical evolution of the spins within each orbital state.
Our model follows an approach previously taken to model the effect of chemical reaction kinetics on NMR spectra~\cite{Knecht2016a}.

\subsubsection{II.1 Mathematical Description of the Model}\label{SI21}

We model the room temperature NV$^-$ center as a system made up of three distinct electronic states: The ground state (gs), the excited state (es), and a meta stable singlet state (s).
We neglect the distinct $E_x$ and $E_y$ orbital branches in the excited state as they are efficiently averaged at room temperature, as well as the existence of two singlet states - we assume that there is only one such singlet state. We also neglect laser induced ionization to the NV$^0$ state.

First we define a separate spin density operator for each orbital state (labelled with $\alpha$) which evolves coherently per the Liouville-von Neumann equation of motion
\begin{equation}
\frac{\text{d}}{\text{d}t}\hat{\rho}_\alpha = \hat{\hat{L}}_\alpha\hat{\rho}_\alpha,
\label{schrödingerequation}
\end{equation}
where the carets denote operators, double carets denote superoperators, and $\alpha$ indexes over the different orbital states.
The commutation superoperator $\hat{\hat{L}}_\alpha$ in Equation\,\eqref{schrödingerequation} can be calculated from the corresponding Hamiltonian $\hat{H}_{\alpha}$ as 
\begin{equation}
\hat{\hat{L}}_\alpha = -i\left(\hat{H}_\alpha\otimes E_\alpha - E_\alpha\otimes\hat{H}_\alpha^{T}\right)\,,
\end{equation}
where $\alpha\in\{\text{es},\text{gs},\text{s}\}$ denotes the orbital, $E_\alpha$ is the identity matrix of the same dimensionality as $H_\alpha$ and $T$ denotes matrix transposition.
The Hamiltonians for each orbital state are given by:
\begin{align}
\hat{H}^{\rm gs, es}/h &= D_0^{\rm gs, es}\hat{S}_z^2 + \hat{\bm{S}}\cdot\bm{A}^{\rm gs, es}\cdot\hat{\bm{I}} + \gamma_{S} \bm{B}\cdot\hat{\bm{S}} + \gamma_{I}\bm{B}\cdot\hat{\bm{I}}\, \\
\hat{H}^{\rm s}/h &= \bm{B}\cdot\hat{\bm{\sigma}}\,. 
\end{align}
where $\hat{\bm{S}}$ and $\hat{\bm{I}}$ are angular momentum operators for the electron and $^{15}$N nucleus acting on the joint electron/nuclear Hilbert space, and $\hat{\bm{\sigma}}$ are the 2x2 spin-1/2 matrices.
The constants and coupling tensors are defined in table\,\ref{NVparams}.
Next, we couple these differential equations with additional (real valued) superoperators corresponding to the incoherent optical pumping process.
These superoperators act to reduce or increase the population of a given spin state and thus take the role of spin-selective relaxation superoperators
\begin{align}
\frac{\text{d}}{\text{d}t}\hat{\rho}_{gs} = \ & \hat{\hat{L}}_{gs}\hat{\rho}_{gs} - k_{\rm green}\hat{\rho}_{gs} + k_{\rm red}\hat{\rho}_{es}
 + k_{s1}(\hat{\hat{S}}^{\otimes}_{+1} + \hat{\hat{S}}^{\otimes}_{-1})\hat{\rho}_{s} + k_{s0}\hat{\hat{S}}^{\otimes}_{0}\hat{\rho}_{s} \nonumber\\
\frac{\text{d}}{\text{d}t}\hat{\rho}_{s}  = \ & \hat{\hat{L}}_{s}\hat{\rho}_{s} - (2k_{s1}+k_{s0})\hat{\rho}_{s}
 + k_{\rm ISC}^{ms=|1|}\hat{\hat{T}}^{e}(\hat{\hat{P}}_{1} +\hat{\hat{P}}_{-1})\hat{\rho}_{es}  + k_{\rm ISC}^{ms=0}\hat{\hat{T}}^{e}\hat{\hat{P}}_{0}\hat{\rho}_{es} \nonumber\\
\frac{\text{d}}{\text{d}t}\hat{\rho}_{es} = \ & \hat{\hat{L}}_{es}\hat{\rho}_{es} + k_{\rm green}\hat{\rho}_{gs} - k_{\rm red}\hat{\rho}_{es}
 - k_{\rm ISC}^{ms=|1|}(\hat{\hat{P}}_{+1} +\hat{\hat{P}}_{-1})\hat{\rho}_{es} - k_{\rm ISC}^{ms=0}\hat{\hat{P}}_{0}\hat{\rho}_{es},
\label{diffeqs}
\end{align}
where $\hat{\hat{P}}_{\pm1,0}$ are projection superoperators that project $\hat{\rho}_{\alpha}$ onto the NV-electron spin state with $m_S=\pm 1,0$ while leaving the dimensionality of $\hat{\rho}_{\alpha}$ unchanged.
This superoperator ensures that the rate of inter system crossing (ISC) out of the excited state depends on the instantaneous spin state population of $\hat{\rho}_{es}$.
Next, $\hat{\hat{T}}^{e}$ is a partial trace superoperator that acts on a 36-dimensional joint electron/nuclear density operator and traces out the NV-electron degrees of freedom, leaving a 4-dimensional density operator corresponding only to $^{15}$N.
Finally, $\hat{\hat{S}}^{\otimes}_{0,\pm1}$ is a direct product superoperator that acts on a 4-dimensional $^{15}$N density operator and turns it into a joint electron/nuclear density operator with the NV-electron in the state $ms=0,\pm1$.
Note that because their effect changes dimensionality of $\hat{\rho_{\alpha}}$, the matrix representations of $\hat{\hat{T}}^{e}$ and $\hat{\hat{S}}^{\otimes}_{0,\pm1}$ are not square.
Further, since none of these superoperators act on the $^{15}$N degrees of freedom this model assumes that the nuclear spin state is preserved throughout the optical pumping process.
Equation\,\eqref{diffeqs} is normalized such that the sum of the traces of the spin density operators $\rho_\alpha$ is equal to 1, meaning $\text{Trace}\{{\hat{\rho}_\alpha}\}$ is the fractional population of the total system in orbital state $\alpha$.
We will give procedures for generating the corresponding matrix representations for the superoperators in Eq.\,\eqref{diffeqs} later in this section.

Figure\,\ref{FigA} shows a pictorial representation of the processes modeled by Eq.\,\eqref{diffeqs}.
The ground and excited states are coupled via optical excitation with rate $k_{\rm green}$ and radiative decay with rate $k_{\rm red}$ respectively. 
The spin selective ISC causes non-radiative transitions from the excited state to the singlet, described by the rates $k_{\rm ISC}^{ms=\pm1,0}$. 
Relaxation from the electronic singlet into the ground state is also electron spin selective, with rates given by $k_{ms=1}$ and $k_{ms=0}$ respectively.
The precise values for the ISC and relaxation rates have been subject to considerable debate. At the end of these section we present modeling results for different sets of these parameters.

\begin{figure}[h]
\includegraphics[width=8.6cm]{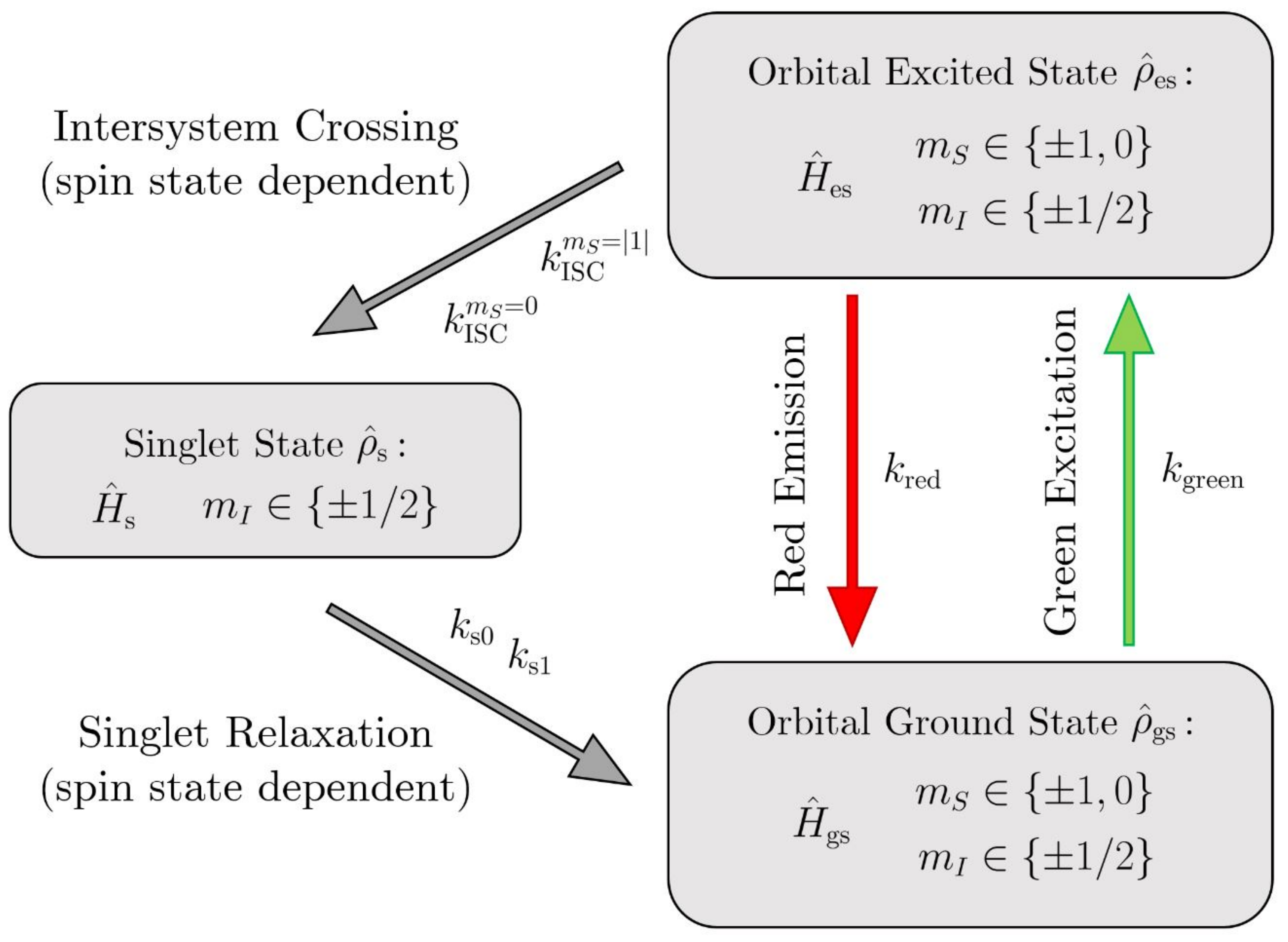}
\caption{\label{FigA} Level Structure for the optical pumping model. Each of the three orbital states is governed by its own Hamiltonian.}
\end{figure}

By concatenating the vector representations of $\hat{\rho}_{gs}$, $\hat{\rho}_{s}$, and $\hat{\rho}_{es}$ into a single 76x1 spin density operator for the entire system, $\hat{\rho}$, Eq.\,\eqref{diffeqs} can conveniently be cast in matrix form:
\begin{gather}\label{matrixeq}
\tiny
\frac{\rm{d}}{{\rm{d}}t}\begin{bmatrix} \hat{\rho}_{gs} & \\ \hat{\rho}_{s} &\\ \hat{\rho}_{es} \end{bmatrix}
=
\begin{bmatrix}
\hat{\hat{L}}_{gs} - k_{\rm green} \hat{\hat{E}}_{36} &+ k_{s\pm1}(\hat{\hat{S}}^{\otimes}_{1} + \hat{\hat{S}}^{\otimes}_{-1}) + k_{s0}\hat{\hat{S}}^{\otimes}_{0} &+ k_{\rm red}\hat{\hat{E}}_{36} \\
0                                                    & \hat{\hat{L}}_{s} - (2k_{s\pm1}+k_{s0})\hat{\hat{E}}_{4}                                    & + k_{ISC}^{m_S=|1|}\hat{\hat{T}}^{e}(\hat{\hat{P}}_{1} +\hat{\hat{P}}_{-1})+k_{ISC}^{m_S=0}\hat{\hat{P}}_{0}\\
k_{\rm green}\hat{\hat{E}}_{36}                         & 0                                                                                                & \hat{\hat{L}}_{es} -k_{f}\hat{\hat{E}}_{36} - k_{ISC}^{m_S=|1|}(\hat{\hat{P}}_{1} +\hat{\hat{P}}_{-1}) - k_{ISC}^{m_S=0}\hat{\hat{P}}_{0}
\end{bmatrix}
\begin{bmatrix} \hat{\rho}_{gs} & \\ \hat{\rho}_{s} &\\ \hat{\rho}_{es} \end{bmatrix},
\end{gather}
where $E_{36(4)}$ is a 36x36(4x4) identity matrix.
In the following, we will call the resulting 76x76 dimensional matrix $\hat{\hat{A}}$.
Note that because of how we formed $\hat{\rho}$ by concatenation it cannot represent coherences between different orbital states, however such coherences are unlikely to be significant and are not necessary to explain the physics of interest in this work.
We also note that since we use identity superoperators to represent the radiative processes, spin-spin coherences will be preserved under optical excitation and radiative decay within this model.
Since the matrix $\hat{\hat{A}}$ commutes with itself for all values of time $t$, Eq.\,\eqref{matrixeq} can easily be integrated and thus the time evolution of the system can be calculated as
\begin{equation}
\hat{\rho}(t) = \mathrm{e}^{2\pi \hat{\hat{A}} t}\hat{\rho}(0),
\label{trho}
\end{equation}
for any time $t$. At any given time the first 36 entries of $\hat{\rho}(t)$ corresponds to $\hat{\rho}_{gs}(t)$ written in vector form, the next 4 entries are $\hat{\rho}_s(t)$ and the last 36 are $\hat{\rho}_{es}(t)$.
We take the predicted instantaneous photoluminescence for $\hat{\rho}(t)$ as the fractional population in the excited electronic state,
\begin{equation}
\text{PL}(t) = \text{Trace}\{\hat{\rho}_{es}(t)\}\,.
\label{PLformula}
\end{equation}
We note that in this work we use Eq.\,\eqref{trho} to calculate the time evolution of the system both in the presence and and absence of green illumination by choosing different values for the parameter $k_{\rm green}$ and calculating the propagation piecewise with different $\hat{\hat{A}}$ matrices.

\subsubsection{II.2 Matrix Representations of Superoperators}\label{SI22}

We define the matrix representations of the $^{15}$N spin operators on the joint electron/nuclear space as $\hat{I}_{x/y/z} = \hat{\sigma}_{x/y/z} \otimes \hat{E}_{3}$ and the NV-electron spin operators as $\hat{S}_{x/y/z} = \hat{E}_{2} \otimes \hat{\lambda}_{x/y/z}$, where $\hat{\lambda}$ are the 3x3 spin-1 matrices. The action of the partial trace superoperator is defined by $\hat{\hat{T}}^e \hat{I}_{x/y/z} = \text{Trace}\{\hat{E}_3\}\hat{\sigma}_{x/y/z}$. In the numerical implementation of our model we vectorize operators as 36x1 or 4x1 dimensional column vectors. The matrix representation of $\hat{\hat{T}}^e$ is determined following the procedure layed out in \cite{Knecht2016a} and reproduced here for completeness:
\begin{equation}
\{\hat{\hat{T}}^e\}_{\alpha\beta} = 
\begin{cases}
\quad 1 &\text{, if } m=n\\
\quad 0 &\text{, else}\\
\end{cases}
\end{equation}
where
\begin{align}
\alpha &= (i-1)\cdot d_I + j \nonumber\\
\beta &= (((i-1)\cdot d_S+m)-1)\cdot d_S \cdot d_I + (j-1)\cdot d_S + n\,,
\end{align}
where $i,n=1\,...d_I$ count through the degrees of freedom of the first subspace (nuclear in our case), and $j,m=1\,...d_S$ count through the degrees of freedom of the second subspace (NV-electron in our case).
The resulting $\hat{\hat{T}}^{e}$  matrix is 4x36 dimensional.

The matrix representation of the direct product superoperator $\hat{\hat{S}}^{\otimes}_{0,\pm1}\hat{E}_2 = \hat{S}_{0,\pm1}$ is
\begin{equation}
\{\hat{\hat{S}}^{\otimes}_{0,\pm1}\}_{\alpha\beta} = \{\hat{\kappa}_{0,\pm1}\}_{mn}
\end{equation}
where
\begin{align}
\alpha &= (((i-1)\cdot d_S+m)-1)\cdot d_S \cdot d_I + (j-1)\cdot d_S + n \nonumber\\
\beta &= (i-1)\cdot d_I + j \,,
\end{align}
and
\begin{equation}\label{ElectronRhoList}
\hat{\kappa}_{+1} = \left( \begin{matrix} 1&0&0\\0&0&0\\0&0&0\\ \end{matrix} \right) \qquad
\hat{\kappa}_0 		= \left( \begin{matrix} 0&0&0\\0&1&0\\0&0&0\\ \end{matrix} \right) \qquad
\hat{\kappa}_{-1} = \left( \begin{matrix} 0&0&0\\0&0&0\\0&0&1\\ \end{matrix} \right) \,.
\end{equation}
The numbers $i$, $j$, $n$, $m$, $d_S$ and $d_I$ are the same as defined above.
The resulting dimensionality of the matrix representation of $\hat{\hat{S}}^{\otimes}_{0,\pm1}$ is 36x4.

Finally, we define the matrix representation for the projection superoperator,
\begin{equation}
\hat{\hat{P}}_{0,\pm1} = \sum_{k=1}^{4} \rho_{0,\pm1}^k \cdot (\rho_{0,\pm1}^k)^T
\label{eq:projection}
\end{equation}
where $\rho_{0,\pm1}^k = \hat{\kappa}_{\rm nuc}^k\otimes \hat{\kappa}_{0,\pm1}$ is the 36x1 column vector representation of the joint 6x6 density matrix operator and 
\begin{equation}
\hat{\kappa}_{\rm nuc}^1 = \left(\begin{matrix} 1&0\\0&0\\ \end{matrix}\right)\qquad
\hat{\kappa}_{\rm nuc}^2 = \left(\begin{matrix} 0&1\\0&0\\ \end{matrix}\right)\qquad
\hat{\kappa}_{\rm nuc}^3 = \left(\begin{matrix} 0&0\\1&0\\ \end{matrix}\right)\qquad
\hat{\kappa}_{\rm nuc}^4 = \left(\begin{matrix} 0&0\\0&1\\ \end{matrix}\right).
\end{equation}
While $\hat{\hat{T}}^{e}$ and $\hat{\hat{S}}^{\otimes}_{0,\pm1}$ both change the dimensionality of the state they operate on, $\hat{\hat{P}}_{0,\pm1}$ is a square 36x36 matrix and thus preserves dimensionality. The sum in equation\,\eqref{eq:projection} ensures that $\hat{\hat{P}}_{0,\pm1}$ projects onto a particular electronic spin state, while leaving the nuclear spin unchanged.

\subsubsection{II.3 Details on Simulation Evaluation}\label{SI23}

In order to simulate the PL of the all-optical pulse sequence introduced in the main text using the model described above, one needs to choose values for the various NV parameters and transition rates $k$. Here, apart from $D_0^{\rm gs}$ which we determine experimentally as described later in the SI, we use literature values for these simulation parameters. The exact numbers employed are listed in table\,\ref{NVparams} and originate from \cite{Gali2009a, Felton2009a}. For the optical transition rates, we consider five different sets of numbers, labeled parameter sets 1 to 5, as shown in the table\,\ref{PUMPparams}.

\begin{center}
\begin{table}[h]
\begin{tabular}{|m{0.75cm}c||c|c|}\hline
$D_0^{\rm gs}$ 				  &$\left[\text{MHz}\right]$		&       $+2870.760402$		\\ \hline		
$A_\parallel^{\rm gs}$  &$\left[\text{MHz}\right]$		&       $+3.03$						\\ \hline	
$A_\perp^{\rm gs}$  	 	&$\left[\text{MHz}\right]$		&       $+3.65$						\\ \hline			
$D_0^{\rm es}$				  &$\left[\text{MHz}\right]$		&       $+1420$						\\ \hline	
$A_\parallel^{\rm es}$  &$\left[\text{MHz}\right]$		&       $-57.8$						\\ \hline	
$A_\perp^{\rm es}$  	 	&$\left[\text{MHz}\right]$		&       $-39.2$						\\ \hline	
$\gamma_S$  	 					&$\left[\text{MHz/G}\right]$	&       $0.000431744$			\\ \hline	
$\gamma_I$			  	 		&$\left[\text{MHz/G}\right]$	&       $2.802494716$			\\ \hline	
\end{tabular}
\caption{\label{NVparams}Numeric values for the employed NV parameters.}
\end{table}
\end{center}

\begin{center}
\begin{table}[h]
\begin{tabular}{|m{1.5cm}c||c|c|c|c|c|}\hline
& & \textbf{ Parameter	}	 &\textbf{ Parameter }		& \textbf{ Parameter } 	& \textbf{ Parameter } 	& \textbf{ Parameter } 	\\ 
& & \textbf{ Set 1	}	 		 &\textbf{ Set 2 }				& \textbf{ Set 3 } 			& \textbf{ Set 4 } 			& \textbf{ Set 5 } 			\\ \hline
$k_{\rm red}$ 				   &$\left[\text{MHz}\right]$	&      66		&      77		&   63.70		&    63.2		&    67.4		 \\ \hline				
$k_{\rm ISC}^{m_S=0}$    &$\left[\text{MHz}\right]$	&       0		&       0		&   12.97		&    10.8		&     9.9		 \\ \hline	
$k_{\rm ISC}^{m_S=|1|}$  &$\left[\text{MHz}\right]$	&      57		&      30		&   80.00		&    60.7		&    96.6		 \\ \hline	
$k_{s0}$ 								 &$\left[\text{MHz}\right]$	&     1.0		&     3.3		&    3.45		&     0.8		&    4.83		 \\ \hline	
$k_{s1}$ 								 &$\left[\text{MHz}\right]$	&     0.7		&       0		&    1.08		&     0.4		&   1.055		 \\ \hline	
Reference &  & \cite{Zhang2021a} & \cite{Manson2006a} & \cite{Robledo2011a} & \cite{Tetienne2012a} & \cite{Gupta2016a} \\ \hline
\end{tabular}
\caption{\label{PUMPparams}Numeric values for of the optical transition rates used in our model. We consider five different parameter sets. In the main text, we use set 4.}
\end{table}
\end{center}

Note that parameters sets 1 and 2 assume $k_{\rm ISC}^{m_S=0}=0$, prohibiting inter system crossing for $m_S=0$ population in the excited state.
Using the values of either of these sets, we fully simulate the all-optical sequence described in the main text, that consists of a green initialization pulse, followed by a interval of free evolution in the dark and subsequent PL readout during a short green laser pulse.
First, we simulate optical pumping of $\hat{\rho}(0)$ at NV saturation for $3000$\,ns, e.g. setting $k_{\rm green} = s\cdot k_{\rm red}$ with $s=35\%$ as determined experimentally on the single NV sample.
To that end, we choose $\hat{\rho}(0)$ such that all six states in the ground electronic orbital are occupied with each $1/6$, and the excited orbital states and the single states are empty.
Afterwards, we run the model with $k_{\rm green}=0$, propagating the optically pumped state in the dark for $290$\,ns.
This empties out the singlet into the ground state. Subsequently, still with $k_{\rm green}=0$, we propagate the system for a duration $\tau$, followed by a readout made up of $350$\,ns with $k_{\rm green}=s\cdot k_{\rm red}$.
This readout pulse is simulated in steps of $1$~ns, and the total PL($\tau$) is the sum of the PL of each step.
Plotting $\text{PL}(\tau)$ yields an oscillation whose frequency $f$ we determine by taking the position of maximum of the corresponding Fourier spectrum.
The contrast $C$ of the oscillation in $\text{PL}(\tau)$ is also determined via the Fourier spectrum: We compute it as $C=4V/V_0$ where $V$ is the amplitude of the detected Fourier space peak, and $V_0$ is the Fourier space peak at zero.

\begin{figure}[h]
\includegraphics[width=8.6cm]{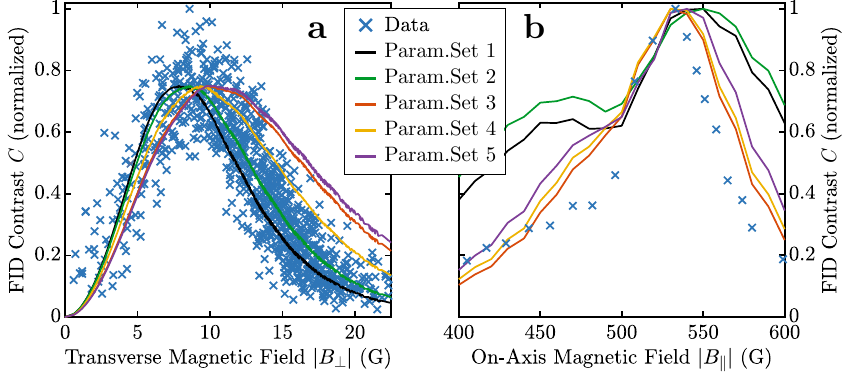}
\caption{\label{FigXmodels} The data shown in Fig\,2b and d in the main text, together with the results of the numerical simulation, run with each of the five optical transition rate parameter sets. Overall, set 4 has the best agreement with the data, which is why model 4 is used for all numerical results shown in the main text.}
\end{figure}

With this procedure, we simulate both the nuclear precession frequency $f$ as well as the precession contrast $C$, for $\bm{B}$-field values corresponding to every pixel shown in Fig\,2b in the main text.
The numeric result for the precession contrast $C$ is shown in Fig.\,\ref{FigXmodels}a, where for ease of comparison, the experimental data has been normalized to 1 and the simulation to 0.75.
The plot demonstrates that the numeric result depends strongly on the chosen optical transition rates.
Specifially, depending on the employed optical set of parameters, we find maximal contrast for $B_\perp$ between $7.9$ and $10.2$\,G.

In Fig\,\ref{FigXmodels}b we show the results of running the model for $\bm{B}$-field values that correspond to the measurement shown in Fig\,2d in the main text.
Again, it is evident that the choice of optical transition rates has a significant effect on the numeric result.
All set of parameters predict a global maximum in $C$ between 530 and 550\,G.
Interestingly, sets 1 and 2 which dictate $k_{\rm ISC}^{m_S=0}=0$ differ drastically from the other parameter sets with $k_{\rm ISC}^{m_S=0}\neq0$ in that they show a second local maximum for $C$ at about 470\,G.
This is not consistent with our data.

Overall, we find that parameter set 4 has the best agreement with our data, and thus we use set 4 for all simulations shown in the main text's figures.
Note that since $\text{PL}(\tau)$ is not true to scale, we normalize all numeric results for $C$ shown in the main text to the corresponding data.

Finally, we note that the numeric model agrees very well with the analytic predictions for the nuclear precession frequency.
This excellent agreement is demonstrated in the main text's figure 3b.
We want to emphasize, however, that not only the precession frequency $f_{\rm nuc} = \gamma_I |\bm{B}_{\rm eff}|$ of numerical and analytical predictions agree well, but also the tilt angle $\vartheta$ of $\bm{B}_{\rm eff}$, as shown in Fig.\,\ref{BeffFigureComp}.
Here, $\bm{B}_{\rm eff}$ and its tilt angle $\vartheta$ are calculated numerically by diagonalization of $\hat{H}_{\rm gs}$, or analytically via Eq.\,(2) in the main text.
Combined, this means that both approaches yield the same $\bm{B}_{\rm eff}$.

\begin{figure}[h]
\includegraphics{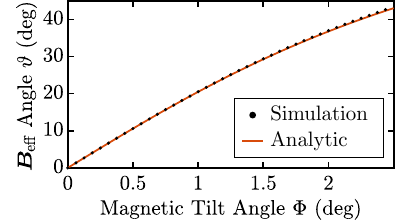}
\caption{Prediction of the tilt angle $\vartheta$ of $\bm{B}_{\rm eff}$ of both our the numeric simulation and the analytic approach based on Van Vleck perturbation theory. These predictions have excellent agreement.}
\label{BeffFigureComp}
\end{figure}

  \subsection{III. Magnetic Field Map}\label{SI3}

In the main text's Fig.\,2b (and 3b), we have shown the same data as in Fig\,1a (and 3a), but plotted against the transverse magnetic field $B_\perp$. Here, we elaborate on how this magnetic field is determined experimentally. 
Since we generate the external magnetic field with a permanent magnet that is mounted on a goniometer, and since this goniometer does not rotate about the NV perfectly, it is in general not accurate to simply measure $|\bm{B}_{\rm ext}|$ and set $B_\perp = \sin(\Phi) |\bm{B}_{\rm ext}|$.
Instead, we measure the magnetic field vector components $B_\perp$ and $B_\parallel$ individually for every other pixel of the main text's Fig.\,2a, using the same experimental conditions, by taking an optically detected magnetic resonance (ODMR) spectrum of each individual pixel.
This way, we experimentally determine all four transition frequencies for a given magnetic field; two corresponding to the hyperfine transitions from the electron spin manifold $\ket{0}$ to the electron spin manifold $\ket{+1}$, and two corresponding to the hyperfine transitions from $\ket{0}$ to $\ket{-1}$.
Then, we determine $D_0^{\rm gs}$ by fitting the ground state Hamiltonian $\hat{H}_{\rm gs}$ as defined in the main text to the four measured transition frequencies of the center pixel, where we enforce $B_\perp=0$ since $\Phi_x=\Phi_y=0$.
We find $D_0^{\rm gs} = 2870.760402$~MHz.
Using this calibration of $D_0^{\rm gs}$, we then determine the magnetic components $B_\parallel(\Phi_x, \Phi_y)$ and $B_\perp(\Phi_x, \Phi_y)$ of all other pixels by fitting $\hat{H}_{\rm gs}$ to the transition frequencies of these pixels, using $D_0^{\rm gs}$ as determined before.
In the end, we interpolate both $B_\parallel(\Phi_x, \Phi_y)$ and $B_\perp(\Phi_x, \Phi_y)$ to twice the density in $\Phi_x$ and $\Phi_y$ to match the pixel density of Fig.\,2a (respectively 3a).
This interpolated result is shown in Fig.\,\ref{FigXmagneticfield}a and b, where $|\bm{B}|=533$~G.
The simulations shown in the main text, as well as the data in Fig.\,2b and 3b are based on these experimentally determined magnetic field values.

\begin{figure}[h]
\includegraphics{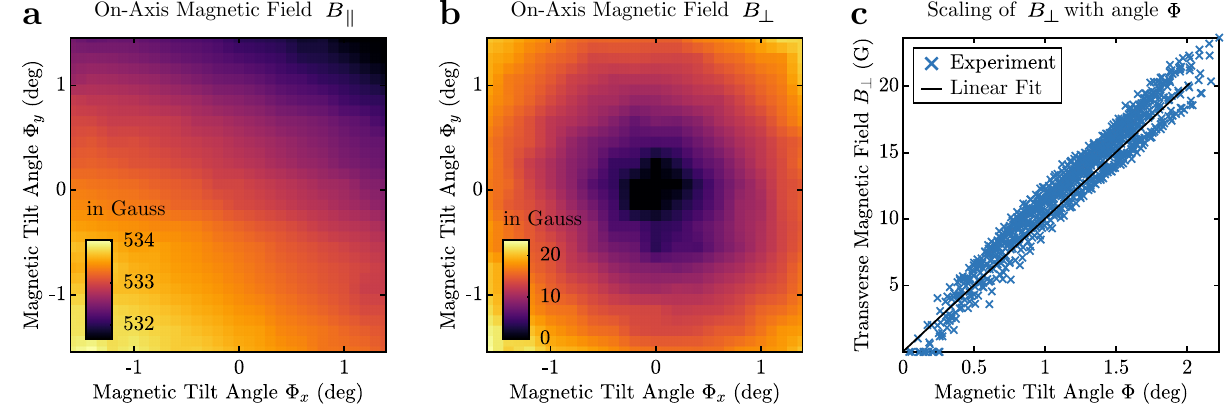}
\caption{\label{FigXmagneticfield} 
\textbf{a} Parallel magnetic field component $B_\parallel$ and \textbf{b} transverse magnetic field component $B_\perp$ as a function of tilt angles $\Phi_x$ and $\Phi_y$, measured via optically detected magnetic resonance of all four microwave-driving $^{15}$N transitions. Each of the shown pixels corresponds to one pixel in the main text's Fig. 2a and 3a. The total magnetic field is about $|\bm{B}|=533$~G.
\textbf{c} Measured transverse magnetic field as a function of total tilt angle $\Phi$, revealing a linear dependance with slope $10.01\,\text{G}/\text{deg}$, allowing for a simple conversion of $B_\perp$ to corresponding $\Phi$ and vice versa.}
\end{figure}

Further, these magnetic field data can be plotted against total angle, $\Phi = (\Phi_x^2 + \Phi_y^2)^{1/2}$, as shown in Fig.\,\ref{FigXmagneticfield}c, revealing a near-linear dependance of $B_\perp$ on $\Phi$. This is to be expected, since for $|\bm{B}|=533$~G and small angles $\Phi$ as investigated here, 
\begin{equation}\label{angleEstimateSlope}
B_\perp = \sin(\Phi\cdot\pi/180^\circ)|\bm{B}|\approx\Phi\cdot\pi/180^\circ|\bm{B}| = \Phi\cdot9.3\,\text{G}/\text{deg}\,.
\end{equation}
The linear fit shown in Fig.\,\ref{FigXmagneticfield}c gives that in our experiment the slope is $10.01\,\text{G}/\text{deg}$. This slightly larger value compared to above prediction (and its not perfeclty linear shape) is explained by the fact that the goniometer employed in our experiments is not rotating the permanent magnet perfectly about the NV, resulting a slight change in total magnetic field whenever the magnet is rotated.

Finally, note that we use the slope of $10.01\,\text{G}/\text{deg}$ to convert $B_\perp$ to $\Phi$ -- but only for data taken at $|\bm{B}|=533$~G because this is the only field where we acquired ODMR data for the entire range of the goniometer. Specifically, as discussed in the main text, at $|\bm{B}|=533$~G the best contrast is achieved at $B_\perp = 9.4$~G which thus corresponds to $\Phi_{\rm opt}=0.94^\circ$.

\begin{figure}[h]
\includegraphics[width=8.6cm]{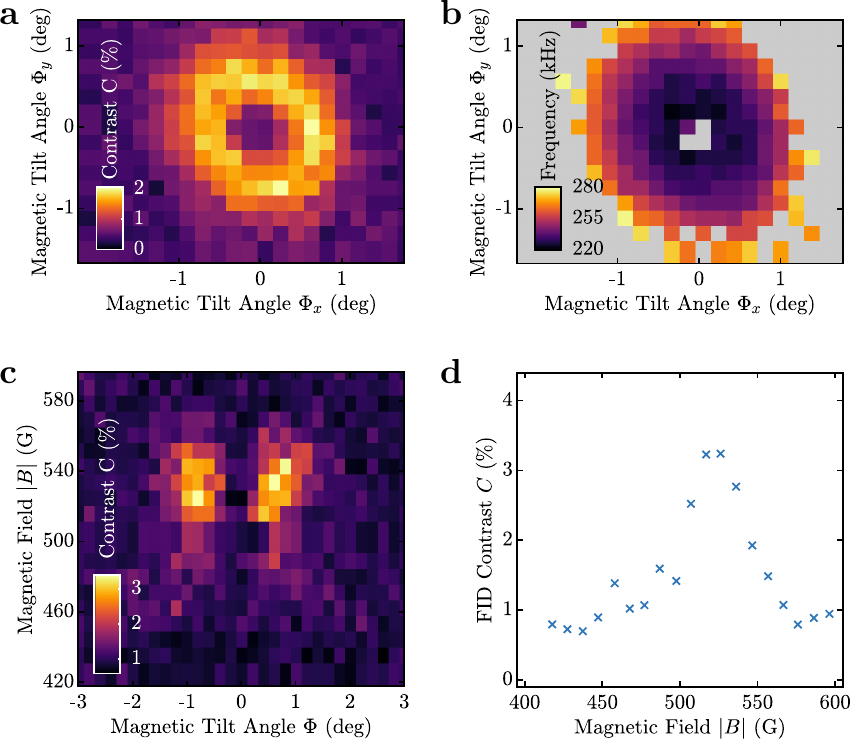}
\caption{\label{FigXensembledonut} Same measurement as shown in the main text in Fig.\,2a, c, d and Fig.\,3a, but here the data were taken on an ensemble of NVs rather than on a single NV. We find the same qualitative and quantitative result, with the only difference being a slightly lower contrast $C$.}
\end{figure}

  \subsection{IV. Ensemble Data}\label{SI4}

In the main text's Fig.\,2 and 3, we discuss the dependance of the of the observed nuclear precession on the applied magnetic field in the context of a single NV.
However, all that data can also be reliably reproduced on a full ensemble of NVs.
In Fig.\,\ref{FigXensembledonut}, we show the results of these ensemble measurements which were taken on the NV ensemble described in the main text: an unstructured, grown diamond with $^{15}$NVs that are preferentially aligned with a specific crystal orientation.
These ensemble measurements reproduce the single-NV data in both qualitative and quantitative way, demonstrating that the all-optical nuclear quantum sensing scheme proposed in this work does scale well to NV ensembles.
We note, however, that the ensemble shows a contrast $C$ of about half the single-NV contrast, which we assign to the minority of non-aligned NVs.

\end{widetext}

\end{document}